\documentclass{elsart}
\usepackage[dvips]{graphicx}

\usepackage{color}
\usepackage{amssymb}
\usepackage{amsmath}
\usepackage{epsf}
\usepackage{subfigure}
\newcommand{\rmd}{\text{d}}
\newcommand{\dfracp}[2]{\dfrac{\partial #1}{\partial #2}}
\newcommand{\dfracd}[2]{\dfrac{\text{d} #1}{\text{d} #2}}
\newcommand{\average}[1]{\left\langle #1 \right\rangle}
\newcommand{\Ucd}{U_{c}^{\ast}}
\newcommand{\norm}[1]{\| #1 \|}

\begin{document}
\begin{frontmatter}
  \title{Stability criteria of the Vlasov equation and
  quasi-stationary states of the HMF model}
  \author{Yoshiyuki Y. Yamaguchi$^{1,3}$}, \author{Julien Barr{\'e}$^2$},
  \author{Freddy Bouchet$^{1,2,4}$}, \author{Thierry Dauxois$^2$} and
  \author{Stefano Ruffo$^1$} \address{1. Dipartimento di Energetica
    ``S. Stecco" and CSDC, Universit{\`a} di Firenze, INFM and INFN, Via S. Marta,
    3 I-50139, Firenze, Italy} \address{2. Laboratoire de Physique,
    UMR-CNRS 5672, ENS Lyon, 46 All\'{e}e d'Italie,\\ 69364 Lyon
    c\'{e}dex 07, France}
  \address{3. Department of Applied Mathematics and Physics,
    Graduate School of Informatics,
    Kyoto University, 606-8501, Kyoto, Japan}
  \address{4. Dipartimento di Fisica Universit{\`a} "La Sapienza" \& INFM
Unit{\`a} di Roma1, P.le A.~Moro 2, I-00185 Roma, Italy}

\thanks[yoshi]{E-mail: yyama@amp.i.kyoto-u.ac.jp}
\thanks[julien]{E-mail: Julien.Barre@ens-lyon.fr}
\thanks[freddy]{E-mail: Freddy.Bouchet@ens-lyon.fr}
\thanks[thierry]{E-mail: Thierry.Dauxois@ens-lyon.fr}
\thanks[stefano]{E-mail: ruffo@avanzi.de.unifi.it}

\begin{abstract}
  We perform a detailed study of the relaxation towards equilibrium in
  the Hamiltonian Mean-Field (HMF) model, a prototype for long-range
  interactions in $N$-particle dynamics.
  In particular, we point out the role played
  by the infinity of stationary states of the
  associated $N\to \infty$ Vlasov dynamics.
  In this context, we derive a new general criterion
  for the stability of any spatially homogeneous distribution,
  and compare its analytical predictions
  with numerical simulations of the Hamiltonian, finite $N$, dynamics.
  We then propose and verify numerically a scenario for
  the relaxation process, relying on the Vlasov equation.
  When starting from a non stationary or a Vlasov 
  unstable stationary initial state,
  the system shows initially a rapid convergence
  towards a stable stationary state of the Vlasov equation
  via non stationary states:
  we characterize numerically this dynamical instability
  in the finite $N$ system by introducing appropriate indicators.
  This first step of the evolution towards Boltzmann-Gibbs equilibrium
  is followed by a slow quasi-stationary process,
  that proceeds through different stable stationary states of
  the Vlasov equation. If the finite $N$ system is initialized
  in a Vlasov stable homogenous state,
  it remains trapped in a quasi-stationary state
  for times that increase with the nontrivial power law $N^{1.7}$.
  Single particle momentum distributions in such a quasi-stationary regime
  do not have power-law tails, and hence
  cannot be fitted by the $q$-exponential
  distributions derived from Tsallis statistics.
\end{abstract}
\begin{keyword}
Hamiltonian dynamics, Long-range interactions, Vlasov equation,
Nonlinear stability, Relaxation times.
\bigskip

{\em PACS numbers:}\\ 05.20.-y Classical statistical mechanics,\\
05.45.-a Nonlinear dynamics and nonlinear dynamical systems,\\
05.70.Ln Nonequilibrium and irreversible thermodynamics,\\
52.65.Ff Fokker-Planck and Vlasov equation.
\end{keyword}
\end{frontmatter}

\section{Introduction}

Relaxation to thermal Boltzmann-Gibbs equilibrium in $N$-particle
Hamiltonian systems with long-range interactions has been recently
the subject of an intense debate~\cite{Springer}. In some cases,
the relaxation time has been shown to be extremely long and to
increase with the number of particles. Hence, the study of these
out-of-equilibrium conditions is of paramount importance for
physical applications. The dynamics of systems with long-range
interactions shows that some states are attained quickly, on time
scales of $O(1)$, and that afterwards the system evolves very
slowly, on time scales diverging with $N$, towards Boltzmann-Gibbs
equilibrium. We call states that evolve on time scales that
diverge with $N$ ``quasi-stationary''. Some of them are
characterized by a particle distribution in the $\mu$-space,
$f(\mathbf{r},\mathbf{p},t) = \sum_i^N \delta (\mathbf{r} -
\mathbf{r}_i(t),\mathbf{p}-\mathbf{p}_i(t))$ (with
$(\mathbf{r}_i,\mathbf{p}_i)$ the position and conjugate momentum
of the $i$-th particle and $\delta$ the Dirac function), which
remains close to a slowly varying smooth distribution for times
that increase with $N$.

It should be remarked that most of the numerical evidences of this
behavior are for $1D$ and $2D$ systems\footnote{Preliminary
indications that it extends to higher dimensions can be found in
Ref.~\cite{giansanti}.}. The theoretical explanation we propose in
this paper, which is developed for mean-field models, extends to
any dimension, as soon as the two body interaction is sufficiently
smooth.  Many recent studies of such quasi-stationary states are
performed for the so-called Hamiltonian Mean Field (HMF) model
(for a review see Ref.~\cite{HMFSpringer}). This model describes
the motion of $N$ rotators under the action of an attractive or
repulsive infinite range cosine interaction. In this paper we will
consider the attractive case. The model then displays a second
order phase transition, which is related to the development of a
dynamical instability of the spatially homogeneous initial state
with gaussian distribution of momenta, at a given value of the
energy~\cite{Inagaki}.

The analysis developed in this paper is based on a theorem due to
Braun and Hepp~\cite{BraunHepp,Spohn}, according to which the
dynamics of a classical $N$-particle system interacting via a two
body and sufficiently regular long-range potential is well
approximated by the associated Vlasov dynamics of the density in
$\mu$-space.  In the case in which the interaction force derives
from a two-body smooth potential, Vlasov equation writes
\begin{equation}
\frac{\partial f}{\partial t}+\mathbf{p}\cdot\nabla_{\mathbf{r}}f
-\nabla_{\mathbf{r}}U \cdot \nabla_{\mathbf{p}}f=0 \quad.
\label{eqboltzmann}
\end{equation}
The mean field macroscopic potential $U$ is a functional of the
probability distribution function $f(\mathbf{r},\mathbf{p},t)$,
which makes the equation nonlinear in $f$. More precisely,
Braun-Hepp's theorem states that, for a mean-field microscopic
two-body smooth potential, the distance \footnote{The distance is
measured in the Wasserstein metric, defined on the space of all
measures.} between two initially close ``weak" solutions of the
Vlasov equation increases at most exponentially in time. The
theorem applies also to singular distributions, e.g. distributions
that have a support on a set of dimension smaller than the one of
the $\mu$-space (for instance, a line in the two dimensional
$\mu$-space of the HMF model). If we apply this result to a large
$N$ particle approximation of a continuous distribution the error
at $t=0$ is typically of order $1/\sqrt{N}$, thus for any ``small"
$\varepsilon$ and any ``large enough" particle number $N$, there
is a time $t$ up to which the dynamics of the original Hamiltonian
and its Vlasov description coincide within an error bounded
by~$\varepsilon$. The theorem implies that this time~$t$ increases
at least as $\ln N$. Extensions to wave particle dynamics of such
a result have also been reported recently~\cite{Firpo,Firpo98}.
Since quasi-stationary states evolve on time scales that diverge
with~$N$, this result suggests that these states might gain their
stability from being ``close'' to some stable stationary states of
the Vlasov dynamics.

Besides the conserved quantities of the Hamiltonian dynamics
(energy, momentum, angular momentum...), the Vlasov description
introduces additional integrals: the so-called Casimirs
\begin{equation}
\label{eq_casimir} C_s[f]=\int s
\left(f(\mathbf{r},\mathbf{p},t)\right) \, d\mathbf{r}
d\mathbf{p}\, ,
\end{equation}
for any function~$s$. These conservation laws are responsible for
the existence of an infinity of stable stationary states for the
Vlasov dynamics. In this paper, we will argue that the existence
of this infinity of stationary states is a possible explanation
for the generic existence of quasi-stationary states in the finite
$N$-dynamics.

This interpretation rises the following questions. May one predict
the quasi-stationary states that emerge after a complex unstable
Vlasov dynamics ?  Among these stationary states, are there some
``statistically preferred'' states ? What governs the relaxation
towards equilibrium of the stationary states, and which is the
scaling of the relaxation time with the number of particles?  We
will address in this paper the first and the third question, while
the issues related to the second question will be only briefly
mentioned.

Before the slow relaxation phase sets in, a fast evolution takes
place on a time scale that is independent of $N$. This phenomenon
is denoted as {\em violent relaxation} in the astrophysical
context~\cite{Lynden68}, and is a consequence of the nonlinear
complex dynamics of the Vlasov equation. After this very rapid
stage, Vlasov dynamics produces thinner and thinner filamentations
of the density $f$, which lead to an apparent equilibrium
described by a coarse grained density function $\bar{f}$. A
statistical mechanics interpretation of this process has been
proposed: for instance, for an initial two level density function,
the equilibrium density is of Fermi-Dirac
type~\cite{ChavanisHouches}. Even if this may be already a partial
answer to our second question above, we will not address in detail
the very delicate point of the convergence of the Vlasov dynamics
towards {\it its} statistical equilibrium in the present paper.

Some authors~\cite{lrt2001} have advanced a challenging
interpretation of the quasi-stationary states, suggesting that
they should be true ``equilibrium" states, obtained by maximizing
the Tsallis~\cite{Tsallis} entropy of the single particle
distribution. We think that a landmark of such an interpretation
would be the assessment of the existence of power law tails in the
single particle momentum (or energy) distributions. Although we
will not enter into a detailed fitting of such distributions using
Tsallis $q$-exponentials, we will present a strong evidence of the
absence of power law tails in the single particle momentum
distribution at any stage of the time evolution, for the whole
class of initial conditions we investigate in this paper, that are
all homogeneous in space (see Section~\ref{sec:long-time}). 
For an initial condition in which the particles are concentrated
in a point and momentum is uniform in an interval around zero, 
the authors of Ref.~\cite{lrt2001} have been able to fit a Tsallis
$q$-exponential to the central part of the momentum distribution,
after imposing an arbitrary cut-off to the tails.
However, these authors do not exhibit any evidence
of existence of power law tails, even for such a special initial state. 

Moreover, as mentioned above, our analysis applies
also to this initial condition although it doesn't correspond to
a stationary state of the Vlasov equation.
At best, Tsallis statistics could
describe the quasi-stationary states obtained from such a special
initial state, but certainly not all of them, in particular those
originated from the large class of homogeneous states that are
studied in this paper.

Moreover, the steadily progressing dynamical evolution observed
for our class of initial conditions does not show any intermediate
``statistically preferred'' state. As stated above, the time
evolution follows a sequence of stationary Vlasov states until,
asymptotically, Boltzmann-Gibbs equilibrium is attained.

It should be however mentioned that
the special initial state studied in~\cite{lrt2001}
is very interesting from a dynamical point of view,
since it shows long-time correlations that are absent
for the homogeneous initial states studied here
(see also ~\cite{Pluchino}). 
Similar initial states produce fractal structures in the 
$\mu$-space for a self-gravitating sheet model ~\cite{koyama}.

As discussed above, Braun-Hepp's theorem suggests that the
similarity between Hamiltonian $N$-particle dynamics and Vlasov
dynamics persists for times that increase as $\ln N$. These times
are  linked with the fastest possible instability of the Vlasov
dynamics. For stable solutions, the appropriate timescale is,
however, the one associated with the fluctuations of the mean
field.  In agreement with this theoretical remark, we present in
Sec.~\ref{sec:long-time} numerical results that indicate that the
persistence of quasi-stationary states is present up to times that
are much longer than Braun-Hepp's $\ln N$. This time scale
increases as a power law in $N$ with a nontrivial exponent.
Similar time scales have been found in gravitational systems.
This is the case of Chandrasekhar's ``collisional" time scale,
which is of order $N/\ln N$~\cite{chandra}. Although such time
scale is similar to those we find in the HMF model, because it
signals the final process of relaxation to Boltzmann-Gibbs
equilibrium, its origin in our model is certainly different.

In order to reformulate the timescale hierarchy sketched above, we
can expect the following scenario:

\begin{enumerate}
\item An initial and fast evolution, well described
  for all initial conditions by the Vlasov dynamics,
  takes place on a timescale independent of the particle number~$N$.

\item The system is then always trapped close to one
  of the numerous stable stationary states of the Vlasov equation.
  This state may be the statistical equilibrium of the Vlasov equation
  (the most probable state with constraints given by the Vlasov
  invariants).

\item The system slowly evolves, on a much longer time scale,
due to ``collisions'', or due to the fluctuations around this
Vlasov stationary state. Consequently this time scale will be a
function of $N$. One can expect that this slow evolution takes
place passing through different stable Vlasov stationary states.

\item Finally, the system reaches a particular Vlasov stable state,
corresponding to the Boltzmann-Gibbs equilibrium state.
\end{enumerate}

We will try to give support to this scenario in the remaining of
the paper. The plan is the following. We first introduce in
Section~\ref{model} the Hamiltonian Mean Field model. We will then
show in Section~\ref{vlasov} that the Vlasov dynamics has an
infinity of stationary states and we propose a new general
stability criterion for any homogeneous distribution (including
non Boltzmann-Gibbs ones). We will compare these analytical
results with numerical simulations of the finite-$N$ Hamiltonian
dynamics. The rapid convergence towards a stable stationary state
of the Vlasov equation is described in
Section~\ref{sec:short-time}. In the case of unstable stationary
states, we show that the exponential destabilization may be
investigated taking advantage of the existence of unstable modes
of the Vlasov dynamics, in accordance with Braun-Hepp's theorem.
The slow evolution towards equilibrium, passing through different
stable Vlasov stationary states is described in
Section~\ref{sec:intermediate-time}.

\section{The Hamiltonian Mean Field model}
\label{model} We will consider the Hamiltonian Mean Field (HMF)
model, whose Hamiltonian is
\begin{equation}
  \label{eq:hamiltonian}
  H_N = \dfrac{1}{2} \sum_{j=1}^{N} p_{j}^{2}
  + \dfrac{1}{2N} \sum_{j,k=1}^{N} [ 1-\cos(\theta_{j}-\theta_{k}) ],
\end{equation}
where $\theta_i\in[-\pi,\pi[$ is the position (angle) of the
$i$-th particle on a circle and $p_i$ the corresponding momentum.
This system can be seen as representing particles moving on a unit
circle interacting via an infinite range attractive cosine
potential, or as classical XY-rotators with infinite range
ferromagnetic couplings~\cite{HMFSpringer}. The magnetization,
defined as
\begin{equation}
  \label{eq:magnetization}
  \overrightarrow{M}(t) = (M_x,M_y) = \dfrac{1}{N} \sum_{j=1}^{N}
  ( \cos\theta_{j},\sin\theta_{j} ),
\end{equation}
or more precisely its modulus, $ M(t) =
||\overrightarrow{M}(t)||\leq 1$, is the main observable that
characterizes the dynamical and thermodynamical state of the
system.

Its introduction allows to write the canonical equations of motion
as follows
\begin{eqnarray}
  \dfrac {d\theta _{j}}{dt}&=&p_{j}\quad,  \nonumber\\
  \dfrac {dp_{j}}{dt}&=&-M_{x}\sin \theta _{j}+M_{y}\cos \theta _{j}\quad.
  \label{eq:canonical}
\end{eqnarray}

Equilibrium statistical mechanics can be derived exactly both in
the canonical and in the microcanonical
ensembles~\cite{antoni-95,barrethesis,bbdrjstatphys}.  In the
ferromagnetic case, that we consider here, a minimal free energy
(maximum entropy) state with a nonvanishing magnetization appears
when lowering temperature (resp. energy per particle) below
$T_c=1/2$ (resp. $U_c=3/4$).  A discontinuity at $T_c$ ($U_c$) in
the second derivative of the free energy (entropy) with respect to
magnetization signals a second order phase transition.  This
transition is between a low energy phase with particles forming a
cluster (rotators pointing towards a preferred direction), and a
high energy phase with particles evenly distributed on the circle
(no preferred direction for rotators).

This theoretical result, valid in the $N\to\infty$ limit, is also
confirmed by direct numerical simulations of the equations of
motion~(\ref{eq:canonical}), which moreover allows a careful
analysis of finite $N$ corrections and give access to the study of
non equilibrium features.  It's in this context that, within the
energy region $U\in[0.5,U_c]$, a class of initial states has been
found that displays an extremely slow relaxation towards
Boltzmann-Gibbs
equilibrium~\cite{lrt2001,antoni-95,LRR98,lrr1999a,lrr1999b,lr2001},
with a relaxation time that increases with $N$. Similar phenomena
occur for other particle systems with long-range interaction
(self-gravitating stars or point vortices \cite{ChavanisHouches}).
Indeed, in the context of two-dimensional
fluid-dynamics~\cite{Arnold_stabilite,Holm_and_co} and plasma
physics~\cite{Holm_and_co}, the existence of an infinity of
stationary states is known since a long time .

These slow relaxation processes have recently attracted
considerable attention, since the HMF model can be considered as a
simple paradigmatic model of long-range interactions, without the
two additional difficulties of gravitational dynamics: singularity
at short range and particle evaporation. As briefly recalled in
the Introduction, Latora \emph{et al.}~\cite{lrt2001} have
carefully analyzed an initial condition where all particles are
located at the same position on the circle (giving initially
$M=1$) and momentum is uniformly distributed over a finite range,
symmetrically around zero. The system shows a fast relaxation
towards a small magnetization state which persists for an
extremely long time, that increases with~$N$. The authors compare
the momentum distribution of such a quasi-stationary state with
Tsallis distributions, obtaining some convincing fit of the
central part of the distribution only after they impose a cut-off
to momentum tails. Montemurro and Zanette~\cite{zanette-02},
analyzing the same initial condition, have even criticized the
existence of a small magnetization plateau in time, by presenting
some numerical evidence that magnetization first evolves towards a
minimum, and then take off again towards the higher equilibrium
value. We will avoid this controversial point by using the
definition of quasi-stationary state given in the Introduction,
i.e. we shall call quasi-stationary a state which still evolves,
but on a time scale that diverges with $N$. Hence, macroscopic
properties are well defined over a sufficiently wide time span to
allow local running time averages, even though the system slowly
and continuously evolves towards equilibrium.

At variance with most previous numerical
experiments~\cite{lrt2001,zanette-02}, we choose here an initial
state where the particles are uniformly distributed on the circle
(hence $M$ is initially close to zero, $M={ O}(1/ \sqrt{N})$) and
momentum has a uniform distribution centered around zero, as
above. We make this choice for two reasons: {\it i}) this state is
a stationary state of the Vlasov equation, that describes the HMF
model in the $N\to\infty$ limit, {\it ii}) the $M=0$ state plays a
relevant role also in the previous numerical
experiments~\cite{lrt2001,zanette-02} and we then thought that it
is better to start directly from it.  Finally, we think that the
analysis of the $M=0$ initial state can also clarify several
aspects of the phenomenology of the $M=1$ initial state used by
the authors above.

As explained in the Introduction, we believe that the Vlasov
description is a useful and appropriate tool to understand the
slow relaxation process. In the following Section, we introduce
and discuss the Vlasov equation for the HMF model.

\section{The Vlasov  dynamics of the HMF Model}
\label{vlasov}

\subsection{Introduction}

In the continuum limit, that is keeping the volume (here the
interval $[-\pi,\pi[$) and the energy per particle fixed as the
number of particles $N\to \infty$, the dynamics governed by
Eqs.~(\ref{eq:canonical}) is described by the Vlasov equation. The
state of the finite $N$ system can be described by a single
particle time-dependent density function
\begin{equation}
 f_d\left(\theta,p,t\right)= \displaystyle \frac{1}{N} \sum_{j=1}^N\delta
\left(\theta -\theta _{j}\left( t\right) ,
  p-p_{j}\left( t\right) \right)
\end{equation}
where $\delta$ is the Dirac function. When $N$ is large, it is
natural to approximate the discrete density $f_d$ by a continuous
one $f\left(\theta,p,t\right)$.  Using this density, also called
$\mu$-space distribution, it is possible to rewrite the two
components of the magnetization~$M$ given by
Eq.~(\ref{eq:magnetization}). They read
\begin{eqnarray}
  \label{eq:magnetization_Vlasov}
  \overline{M}_x\left[ f\right] &\equiv&
  \int f(\theta ,p,t)\, \cos \theta\ \rmd \theta \rmd p\quad,\\
  \overline{M}_y\left[ f\right] &\equiv &
  \int f(\theta ,p,t)\,\sin \theta \ \rmd \theta \rmd p\quad.
  \label{eq:magnetization_Vlasovb}
\end{eqnarray}
Within this approximation, one can write the potential that
affects all the particles as
\begin{equation}
 V\left( \theta \right) \left[ f\right] =
-\overline{M}_x\left[ f\right]\, \cos \theta -
\overline{M}_y\left[ f\right]\,  \sin \theta\quad. \end{equation}
This potential enters the expression of the Vlasov equation
\begin{equation}  \label{eq:vlasov}
  \dfracp {f}{t} + p \dfracp {f}{\theta }
  -\dfracd{V}{\theta} \left[ f\right] \dfracp{f}{p} = 0\quad,
\end{equation}
which governs the spatiotemporal evolution of the density $f$. In
the remaining of the paper, we will omit the over-bar
on $ M_{x} $ and $ M_{y} $ for the sake of simplicity.\\

It is important to notice that the discrete distribution $f_d$, a
sum of Dirac peaks, contains exactly the true dynamics of the
system and is also a solution of the Vlasov
equation~(\ref{eq:vlasov}). Then, introducing a suitably defined
distance on the space of probability measures on
$[-\pi,\pi[\times\mathbb{R}$, it is possible to
show~\cite{BraunHepp,Spohn,Firpo} that the distance between two
solutions $f_1$ and~$f_2$ of Eq.~(\ref{eq:vlasov}) grows at most
exponentially in time
\begin{equation}d\left(f_1(t),f_2(t)\right)\leq
d\left(f_1(0),f_2(0)\right)\,\displaystyle e^{\displaystyle \gamma
t} \quad,
\end{equation}
with a growth rate $\gamma$ which is independent of the initial
conditions. This result, which is the essence of Braun-Hepp's
theorem, heavily relies on the mean field character of the
underlying Hamiltonian dynamics and on the genericity of
exponential instability of trajectories.  Choosing then
$f_1(\theta,p,t)=f_d(\theta,p,t)$ and taking for $f_2(\theta,p,t)$
a continuous approximation of $f_d$, one immediately obtains that
$d\left(f_1(0),f_2(0)\right)\to 0$ as $N$ grows. As a consequence,
the previous result implies that the discrete particle dynamics
converges to Vlasov dynamics when $N\to\infty$, uniformly over all
fixed time intervals $[0,\,T]$.  However, for all fixed $N$ there
is a typical time $\tau \sim \gamma^{-1}$ over which the two
dynamics diverge.

From this analysis, it is therefore natural to expect that
particle and Vlasov dynamics coincide during a time that diverges
as $\ln N$, if $d(f_{d}(0),f(0))\sim 1/N$ and $\gamma$ is
independent of $N$. However, for initial conditions corresponding
to stable stationary solutions of the Vlasov equation, this time
may be much longer, actually of order $N$~\cite{freddy}.

All this explains why the properties of the Vlasov equation are of
particular interest for the study of the particle dynamics. In the
next subsection, we will study the Vlasov dynamics of the HMF
model and its stationary states, with the aim of getting useful
insights on particle dynamics.

\subsection{The Vlasov dynamics and its stationary states}

The Vlasov equation inherits from the particle dynamics the
conservation of the energy $ { H_V} $
\begin{equation}
  \label{eq:Energy}
  { H_V}\left[ f\right] = \int \frac{p^{2}}{2}\, f(\theta ,p,t)\,\rmd \theta \rmd p
  + \frac{1}{2} - \frac{M_{x}^{2}+M^{2}_{y}}{2}
\end{equation}
 and of the total momentum
\begin{equation}
  \label{eq:Impulsion}
  P\left[ f\right]=\int p\,f(\theta ,p,t)\,\rmd \theta \rmd p\quad.
\end{equation}
However, the Casimir's functionals~(\ref{eq_casimir}), defined for
any continuous function $s$, yield an infinity of additional
conserved quantities, linked with the labeling symmetry when
following a fluid particle in the $ \mu $-space. These new
conserved quantities play of course a major role in the dynamics
of the Vlasov equation and thus of the particle dynamics.

For a non stationary initial distribution, the dynamics of the
Vlasov equation is known to give rise to a very complex nonlinear
evolution, characterized by stretching and folding of the initial
distribution, the details of this evolution being usually
unpredictable. However, one may predict the final evolution using
statistical mechanics arguments, in the spirit of the statistical
mechanics of two dimensional conservative
flows~\cite{Robert_Sommeria,Miller} or of the Vlasov-Poisson
equation~\cite{Lynden_Bell,Chavanis}.  One ends up with the most
probable coarse-grained distribution $\overline{f}$, which takes
into account the dynamical invariants. Unfortunately, the
dynamical mixing of the distribution is likely to be incomplete:
the reason lies in the existence of infinitely many stable
stationary solutions of the Vlasov equation, in the neighborhood
of which the system may be trapped. Therefore, Vlasov dynamics
quickly converges (weak convergence) to a stable stationary state
which should be studied in detail.

Equations~(\ref{eq:magnetization_Vlasov})
and~(\ref{eq:magnetization_Vlasovb}) show that the magnetization
$\overrightarrow{M} $ is constant for a stationary
solution~$f(\theta,p)$. This implies that the potential $V$ is
constant. The equation for the stationary states of the Vlasov
equation may thus be considered as a linear first order partial
differential equation. Solutions are then given by densities $f$
that are constant on the characteristics of the equation,
corresponding to the level sets of the individual particle energy
\begin{equation}
  \label{eq:individual-energy}
  e(\theta ,p) = \dfrac {p^{2}}{2}+V\left( \theta \right)
  = \dfrac{p^{2}}{2} 
- M_{x}\cos\theta - M_{y}\sin\theta \quad,
\end{equation}
which corresponds to the energy of a pendulum. It is important to
observe that the total energy $ {H_V} $ given by
Eq.~(\ref{eq:Energy}), is different from the sum of the individual
energies~(\ref{eq:individual-energy}).

Smooth stationary solutions of the Vlasov equation are thus given
by
\begin{equation}
\label{statstateVlasov}
 f\left( \theta ,p\right) =\Phi \left( e\left( \theta ,p\right)
 \right)\quad,\end{equation}
where $\Phi$ is any real function. Moreover, the values of
$M_{x}$, $M_{y}$ and the function $\Phi$ must be  self-consistent.
The Boltzmann-Gibbs equilibrium density
\begin{equation} f_{eq}\left( \theta ,p\right) = A\exp \left(
  -\beta e\left( \theta ,p\right) \right)  \label{eqstructure}\end{equation}
is a particular case, although very important. One may also prove
that stationary states of the Vlasov equation in a moving frame
with constant velocity  $ v $ are given by $ f\left( \theta
,p\right) = \Phi \left( e\left( \theta ,p\right) +vp\right)$.

Let us note that the function~$\Phi$ may be multi-valued for
individual energies~$e$ greater than the energy of the separatrix
(one branch for particles with positive momentum and the other for
negative momenta). We will assume that this does not happen in the
remaining of the paper, for the sake of simplicity (a
generalization would be straightforward).

\subsection{Stability of stationary states of the Vlasov equation}
\label{sec:stability}

As discussed above, the stationary states of the Vlasov equation
are not true stationary states of the particle dynamics. If they
are stable, however, they may explain the long lifetime of
quasi-stationary states in the particle dynamics. Linear stability
results for the stationary states of the Vlasov equation have been
already reported in the case of spatially homogeneous
distributions for both Gaussian~\cite{Inagaki} and  water
bag~\cite{antoni-95} momentum distributions.  In this Section, we
will show that it is possible to derive stability results for
arbitrary spatially homogeneous stationary states, using a method
developed in the context of two dimensional fluid dynamics and in
plasma physics~\cite{Holm_and_co}, based on original ideas
introduced by Arnold~\cite{Arnold_stabilite}.

The authors of Ref.~\cite{Holm_and_co} actually study
\emph{nonlinear stability}, a concept that we would like to
briefly distinguish from other stability concepts.  For a generic
dynamical system, any extremum $f_{0}$ of a conserved quantity $
F\left[ f\right] $ is a stationary point of the dynamics.  It is
said to be \emph{formally stable} if the second variations $
\Delta_2F\left[ \delta f_{1},\delta f_{2}\right]$ of $F$ is
positive definite ($f_0$ is then a minimum) or negative definite
($f_0$ is then a maximum). In the case of the linearized dynamics
around a formally stable point $ f_{0} $, as the second variations
of $ F $ at $ f_{0} $ are conserved, a small perturbation of $
f_{0} $ remains bounded in the norm provided by the second
variations: this state is \emph{linearly stable.} Since this
implies that the spectrum of the linearized dynamics does not have
any negative value, the system is also \emph{spectrally stable.}
It is however not true in general that spectral stability implies
linear stability, and that linear stability implies formal
stability. Finally, nonlinear stability corresponds to the case
where a small perturbation, evolving according to the real
dynamics, remains bounded in some norm. It can be shown that
nonlinear stability implies spectral stability, the converse being
wrong in general, whereas formal stability implies nonlinear
stability only in finite dimensional systems.

In this Section we prove that any stationary state of the Vlasov
equation, defined by Eq.~(\ref{statstateVlasov}) with $ \Phi $
strictly decreasing, corresponds to a critical point of some
invariant functional. Computing the second variations of this
functional, we can therefore exhibit a {\em necessary and
sufficient condition of formal stability} for such a stationary
state.

Let us consider the maximum of the functional
\begin{equation}
F\left[ f\right] = C_{s}\left[ f\right] -\beta H_V\left[ f\right]
-\mu \int f(\theta ,p,t)\, \rmd \theta \rmd p \quad ,
\end{equation}
where $ H_V $ is the energy (\ref{eq:Energy}), $ C_{s} $ is a
Casimir functional (\ref{eq_casimir}) corresponding to a strictly
concave function $s$ and $ \beta$ is positive. Performing the
first variations of this functional, we obtain the equation
\begin{equation}
  s' \left( f_{0}\right) =
  \beta \left( \frac{p^{2}}{2} - \int \cos \left( \theta -\alpha \right)
    f_{0}(\alpha ,p,t)\,\rmd \alpha \rmd p\right)
  + \mu =\beta
 e\left( \theta ,p\right) 
 +\mu \quad ,
  \label{derivs}
\end{equation}
which defines the critical points $ f_{0} $. This yields
\[
f_{0}\left( \theta ,p\right)
=\Psi \left( \beta\, 
e\left( \theta ,p\right)
  +\mu \right)~,
\]
with $ e $ given by~(\ref{eq:individual-energy}) and $\Psi$ the
inverse function of $ s'  $, the derivative of $ s $. The
condition that $ s $ is strictly concave is equivalent to the fact
that $\Psi$ is strictly decreasing.

The computation of the second variation of $ F $ gives:
\begin{equation}
  \label{eq:second_variation}
  \Delta _{2}F\left[ \delta f,\delta f\right]
  = \int s''\left( f_{0} \left( \theta ,p\right) \right)
  \left[ \delta f\left( \theta ,p\right) \right] ^{2} \rmd \theta \rmd p
  + \beta \left( \left( M_{x}\left[ \delta f\right] \right) ^{2}
    + \left( M_{y}\left[ \delta f\right] \right) ^{2}\right)~,
\end{equation}
where $ M_{x} $ and $ M_{y} $ are given by
Eqs.~(\ref{eq:magnetization_Vlasov})
and~(\ref{eq:magnetization_Vlasovb}). As $s$ is strictly concave,
$s''$ is negative.  The first term of $ \Delta _{2}F\left[ \delta
f,\delta f\right] $ is thus clearly negative whereas the second
one is positive. We will consider now only homogeneous states,
corresponding to $ M_{x}=M_{y}=0 $ and $ f_{0}\left( \theta
,p\right) =f_{0}\left( p\right) =\Psi \left( \beta p^{2}/2+\mu
\right) $.

Let us introduce the Fourier series of the perturbation
\begin{equation}\delta f\left( \theta ,p,t\right) = \sum_{n}c_{n}\left( p,t\right)\,
\cos n\theta  + s_{n}\left( p,t\right)\, \sin  n\theta\quad.
\label{dvlpfourierdeltaf}
\end{equation}  From Eq.~(\ref{eq:second_variation}), after
integration on the spatial variable $ \theta $, one obtains the
second variations
\begin{equation}
  \label{eq:second_variation_Fourier}
  \Delta _{2}F\left[ \delta f,\delta f\right]
  = \int s^{\prime \prime }\left( f_{0}\left( p\right) \right)
  \sum_{n>1}\left( c^{2}_{n}\left( p\right)
    + s^{2}_{n}\left( p\right) \right) \rmd p
  + 2G\left( c_{1}\left( p\right) \right)
  + 2G\left( s_{1}\left( p\right) \right)~,
\end{equation}
where we have introduced
\begin{equation}
  G\left( c\left( p\right) \right) \equiv
  \int s^{\prime \prime } \left( f_{0}\left( p\right) \right) c^{2}
  \left( p\right) \rmd p
  + \frac{\beta }{2}\left( \int c\left( p\right) \rmd p\right) ^{2}\quad .
\end{equation}
The terms involving $ c_{n} $ and $ s_{n} $, $ n>1 $, in
Eq.~(\ref{eq:second_variation_Fourier}), are clearly negative
definite, since $ s^{\prime \prime } $ is strictly negative.
Consequently, the second variations of $ F $ are negative definite
if and only if $ G $ is negative definite.

The sign of the function $ G $ can be studied using the
Cauchy-Schwartz inequality:
\begin{eqnarray}
  \left( \displaystyle\int c\left( p\right) \rmd p\right) ^{2}
  &=& \left( \displaystyle \int \displaystyle
    \frac{c\left( p\right) \sqrt{-s^{\prime \prime }
        \left( f_{0}\left( p\right) \right) }}{\sqrt{-s^{\prime \prime }
        \left( f_{0}\left( p\right) \right) }}\rmd p\right) ^{2} \\
  &\leq& \left( \int s^{\prime \prime }
    \left( f_{0}\left( p\right) \right) c^{2}\left( p\right) \rmd p\right)
  \left( \int \frac{\rmd p}{s^{\prime \prime }
      \left( f_{0}\left( p\right) \right) }\right) .
\end{eqnarray}
This inequality leads therefore to
\begin{equation}
  G\left( c\left( p\right) \right)
  \leq \int\limits s^{\prime \prime } \left( f_{0}\left( p\right) \right) c^{2}
  \left( p\right) \rmd p\,
  \left( 1+\frac{\beta }{2}\left( \int \frac{1}{s^{\prime \prime }
        \left( f_{0}\left( p\right) \right) }\rmd p\right) \right)\quad.
\end{equation}
Recalling that $ s^{\prime \prime } $ is strictly negative, we
conclude that if the quantity
\begin{equation}   1+\frac{\beta}{2} \int \frac{\rmd p}{s^{\prime \prime }
\left( f_{0}\left( p\right) \right)} \label{criterai}
\end{equation}
is positive, the function $ G $ is negative. Conversely, when it
is negative, considering the particular function $ c\left(
p\right) =-1/s^{\prime \prime} \left( f_{0}\left( p\right) \right)
$ demonstrates that $ G $ may be positive.

Differentiating with respect to the variable $p$ the equality
\begin{equation} s^{\prime }\left( f_{0}\left( p\right) \right) =\beta\,
  \frac{p^{2}}{2}+\mu~,
\end{equation}
obtained from Eq.~(\ref{derivs}) in the case of homogeneous
states, yields:
\begin{equation} \frac{\beta }{s^{\prime \prime }\left(
    f_{0}\left( p\right) \right)} = \frac{f^{\prime }_{0} \left( p\right)}{ p}
  \quad.
\end{equation}
As $f'_{0}(0)=0$, the  quantity~(\ref{criterai})  can   be written
as
\begin{equation}
  \label{eq:testf}
  I[f_{0}] = 1 + \dfrac {1}{2}
  \int ^{+\infty }_{-\infty } \dfrac {f'_{0}(p)}{p}\rmd p\quad.
\end{equation} This leads to the following equivalence:
\begin{equation}
  \label{eq:theorem}
  f_{0}\text{ is formally  stable}\Longleftrightarrow I[f_{0}] > 0\quad .
\end{equation}
This condition will of course be an excellent criterion to predict
the stability of several initial conditions, and therefore the
lifetime of the corresponding quasi-stationary states. This is
what we will consider now.

Let us note that Inagaki and Konishi \cite{inagakikonishi} have
found a dispersion relation for the linearized Vlasov equation
which leads to the above stability criterion. Hence, in this
particular case, linear stability and formal stability criteria
coincide.

\subsection{Applications of the nonlinear stability criteria}
\label{Applicationsnonlinearstability}

\paragraph{Waterbag distribution}

One of the most widely used initial condition in previous
numerical studies of the HMF model is the so called waterbag
distribution
\begin{equation}
  \label{eq:waterbag}
  f_{\text{wb}}(p) = \left\{
    \begin{array}{ll}
      0 & \quad \text{if } |p|> \bar{p} \\
      1/(2\bar{p}) & \quad \text{if }  -\bar{p}<p< \bar{p}~.
    \end{array}
  \right.
\end{equation}
If $M_x=M_y=0$ (homogeneous state), the relation between energy
$H_V=U$ and $\bar{p}$ is $\bar{p}=\sqrt{6U-3}$. Computing the
first derivative of $f_{\text{wb}}$
\begin{equation}
\label{eq:waterbag_deri}
  f'_{\text {wb}}(p)
  =  \frac{\displaystyle 1}{\displaystyle 2\bar{p}}\left[ \delta
    (p+\bar{p}) -\delta (p-\bar{p}) \right]\quad,
\end{equation}
one obtains the following expression
\begin{equation}
  I[f_{\text {wb}}]=1-\dfrac {1}{2}\dfrac {1}{\bar{p}^{2}}\quad,
\end{equation}
exhibiting that the critical width of the distribution, above
which the waterbag is formally unstable, is
$\bar{p}_{c}=1/\sqrt{2}$. This corresponds to the critical energy
density
\begin{equation}
  U_{c}^{\ast }=\dfrac {7}{12}\quad ,\label{criticalwb}
\end{equation}
as reported earlier~\cite{antoni-95}.

\paragraph{Gaussian distribution} As a second example,
let us consider a Gaussian distribution,
\begin{equation}
  f_{\text {g}}(p)= \sqrt{\frac{\beta}{2\pi  }}\ e^{-\beta p^{2}/2}\quad,
\end{equation}
characteristic of an equilibrium canonical distribution. In this
case, the quantity~(\ref{eq:testf}) can be easily reduced to
\begin{equation}
  I[f_{\text {g}}]=1-\dfrac {\beta}{2} \quad,
\end{equation}
emphasizing that the critical inverse temperature is $ \beta
_{c}^{\ast }=2$  and consequently the critical energy density $
U_{c}^{\ast }=3/4 $. This coincides with the critical point of the
second order phase transition $U_{c}$.

\paragraph{Mixed distribution} Finally let us consider a more general
distribution, namely a mixed distribution between $ f_{\text {wb}}
$ and $ f_{\text {g}} $, defined as
\begin{equation}
  \label{eq:fa}
  f_{a}(p) = (1-a) f_{\text {wb}}(p) + a f_{\text {g}}(p)\quad.
\end{equation}
Thanks to the linearity of the quantity~(\ref{eq:testf}) with
respect to the distribution, the critical energy density for this
mixed distribution $ f_{a} $ is obtained as a linear combination
of both previous results:

\begin{equation}
  U_{c}^{\ast }(a)=\dfrac {7}{12}(1-a) + \dfrac {3}{4}a=
  \dfrac {7}{12}+\dfrac {a}{6}\quad.\label{critmixed}
\end{equation}

Such a result allows to define a phase diagram of the dynamical
critical energy that we will be able to confirm using numerical
simulations. We aim in the following at showing how these
considerations about the Vlasov equation and its stationary states
are useful to understand the dynamics of the discrete particle
system. We begin in Section~\ref{sec:short-time} by studying the
short time evolution, and turn in
Section~\ref{sec:intermediate-time} to the intermediate and long
time behaviour.

\section{Short time behavior}
\label{sec:short-time}

\subsection{The numerical setup}
\label{sec:numerical-setup}

We have numerically integrated the canonical equations of
motion~(\ref{eq:canonical}), by using symplectic 4th-order
integrators, the McLachlan-Atela's~\cite{mclachlan-92} or
Yoshida's~\cite{yoshida} algorithms.  The timestep $dt=0.1$ was
chosen to obtain an energy conservation with a relative accuracy
$|\Delta E/E|$ better than $10^{-7}$.  We will consider initial
conditions with uniform distribution with respect to $\theta$ as
explained in Section~\ref{sec:stability}.  The magnetization being
consequently zero, these states are stationary solutions of the
Vlasov equation~(\ref{eq:vlasov}).

However, the numerical calculations  correspond to simulations
with a {\em finite} number of degrees of freedom, and these
initial states are not anymore stationary a priori. The spatial
coordinates $\theta_{j}$ are randomly chosen in the interval
$[-\pi,\pi[$, which leads to a magnetization of order
$1/\sqrt{N}$ initially. The momenta $p_{j}$ are also randomly
chosen from the given distribution $f(p)$ satisfying the
conditions for the energy
\begin{equation}
  \frac{1}{N}\sum_{j=1}^{N} \dfrac{p_{j}^{2}}{2} = U - \dfrac{1}{2}\quad,
\end{equation}
whereas the conserved total momentum is set to zero
\begin{equation}
  \sum_{j=1}^{N} p_{j} = 0\quad.
\end{equation}
We will present how the numerical results allow us to detect the
critical energy for dynamical instability, using first the most
widely used initial conditions, the waterbag distribution, and
then the mixed ones.

\subsection{Waterbag initial distributions}
\label{sec:waterbag}

\subsubsection{The first peak of the magnetization}
\label{sec:first-peak} 
\begin{figure}[htbp]
  \centering
  \includegraphics[width=7.0cm]{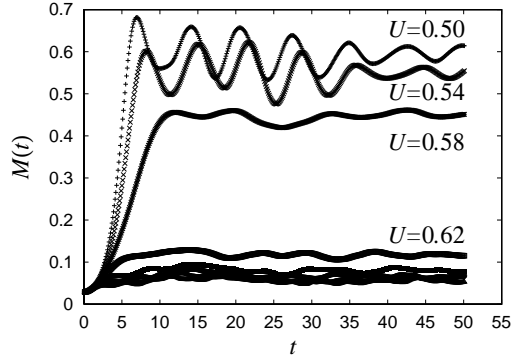}
  \caption{Temporal evolution of $M(t)$ for different values
    of the energy density $U$, when the number of particles is
    $N=10^3$.
    The values of $U$ are from $0.50$ to $0.78$
    with $0.04$ step size.
    The curves correspond to averages over 100 samples.}
  \label{fig:short-time-M}
\end{figure}

\begin{figure}[htbp]
  \centering
    \includegraphics[width=7cm]{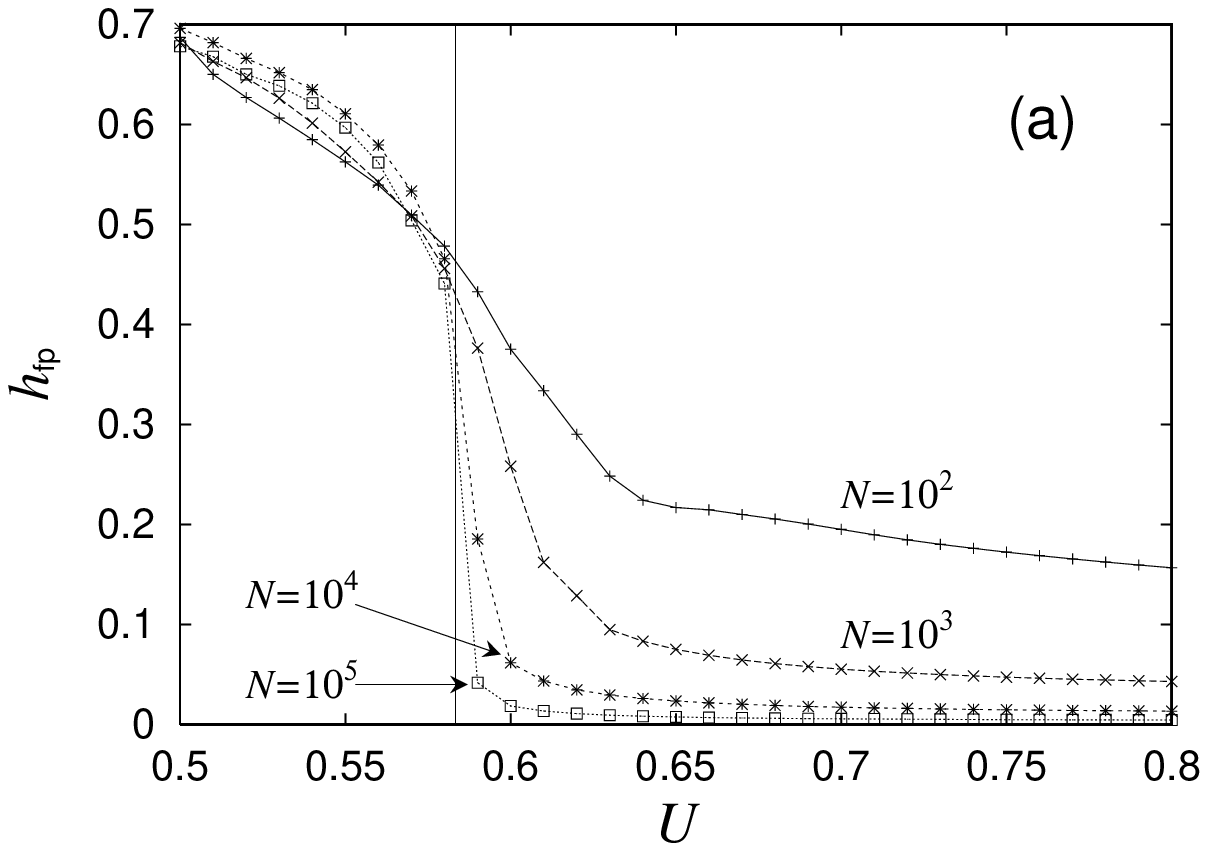}
    \includegraphics[width=7cm]{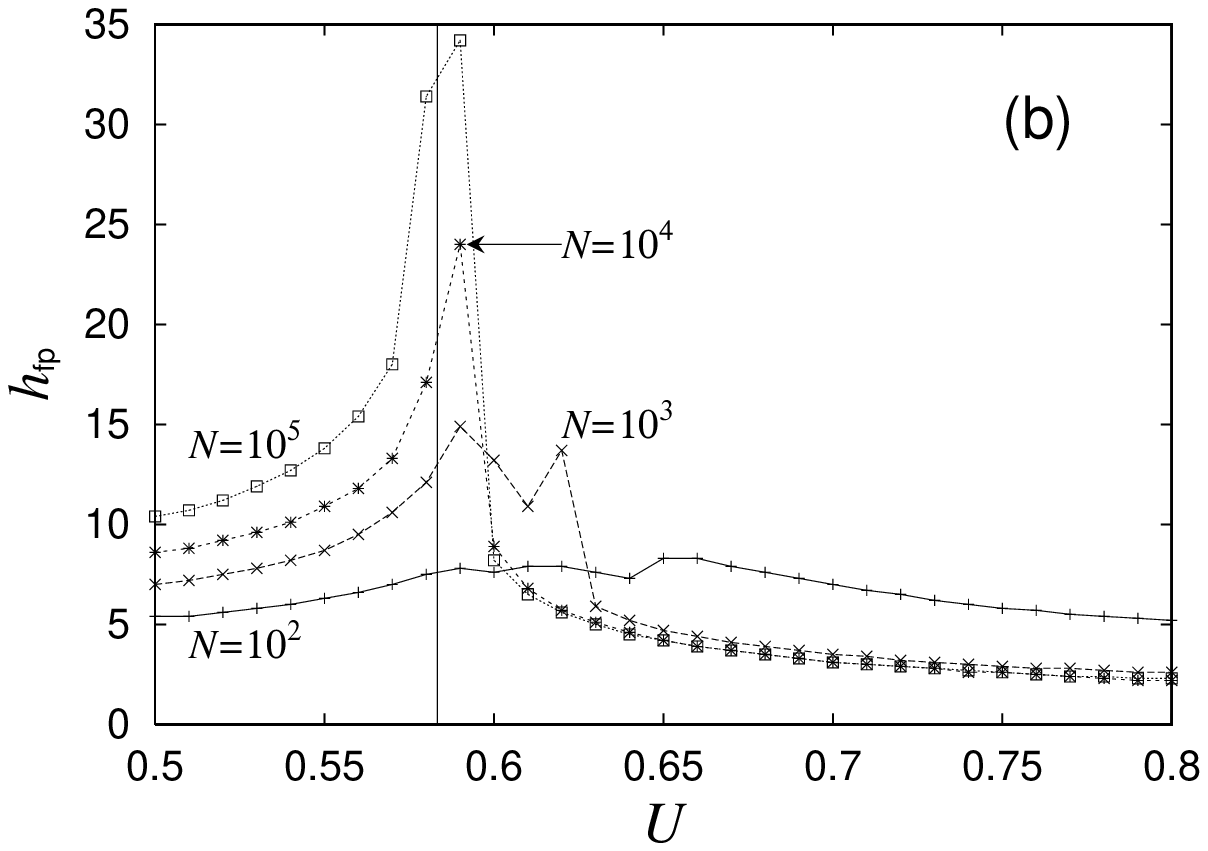}
  \caption{
    The first peak height (a) and the first peak time (b)
    vs. energy density~$U$.
    The curves correspond to different particle numbers
    $N=10^2,10^3,10^4$ and $10^5$
    from top to bottom in $U>\Ucd$ for panel (a)
    and from bottom to top in $U<\Ucd$ for panel (b),
    keeping in mind that every curve is obtained
    averaging over $100$ (resp. $20$) samples
    for $N\leq 10^4$ (resp. $N=10^5$).
    The vertical line represents the theoretical critical energy density
    $U_{c}^{\ast}$, given by Eq.~(\ref{criticalwb}).
  \label{fig:first-peak-a0.0}}
\end{figure}

Figure~\ref{fig:short-time-M} presents the
initial temporal evolution of the magnetization $M(t)$ for
different values of the energy density, but with the same number
of particles, for a system initialized with a waterbag
distribution in momentum and a homogeneous one for the angles($M$
is close to zero since initially $M={ O}(1/ \sqrt{N})$). Averages
over a set of initial conditions (samples) are taken.  These
results (already partially reported elsewhere~\cite{Yoshipaper})
show that the initial time evolution of the magnetization,
starting from such a non equilibrium initial condition, is quite
different depending on whether $U$ is larger or smaller than $\Ucd
\approx 0.583$.  If $U>\Ucd$, but still below the second order
phase transition energy $U_c$, the magnetization remains close to
the $M=0$ initial value and does not show any tendency towards the
nonvanishing equilibrium value (the long time relaxation to
equilibrium will be discussed in Section~\ref{sec:long-time}). For
$U<\Ucd$, instead, the magnetization shows a fast relaxation to a
non vanishing value which is close to equilibrium. Relaxation
proceeds through repeated oscillations that damp after a
relatively short time. In order to characterize quantitatively
this behavior, let us focus on the first peak of $M(t)$, by
studying its height and its time of occurrence as a function of
the energy density~$U$, as presented in
Fig.~\ref{fig:first-peak-a0.0} for increasing particle numbers.

Figure~\ref{fig:first-peak-a0.0}(a) emphasizes that the first peak
height vanishes in the energy region above $\Ucd$ as $N$
increases, in agreement with the theoretical predictions of the
Vlasov equation, which imply that the $M=0$ state becomes stable
in this energy range in the continuum ($N\to\infty$) limit.  The
initial fluctuations of the magnetization, which are of order
$1/\sqrt{N}$, do not grow when $U>\Ucd$, because the waterbag
state is Vlasov stable.  On the contrary, these fluctuations
rapidly grow for $U<\Ucd$, leading in a short time to a
nonvanishing magnetization state which is close to equilibrium.

Similar indications come from the first peak time, shown in
Fig.~\ref{fig:first-peak-a0.0}(b). For $U>\Ucd$ this quantity
clearly shows a convergence to an asymptotic value as $N$
increases, but this value sharply increases as one approaches
$\Ucd$ from above, signaling the instability of the $M=0$ initial
state. The behavior below $\Ucd$ is less clear since the first
peak time does not yet show the saturation expected on the basis
of the Vlasov equation for the values of $N$ considered here. This
might perhaps be due to the limited validity in time of the Vlasov
description when starting from an initially unstable state. We
will however show in the next Section that at least the early
exponential instability is well reproduced.

Both quantities, the first peak height and the first peak time,
are therefore useful indicators of the presence of a dynamical
critical energy $\Ucd$, as predicted theoretically in the
continuum limit. To summarize, we could say that on a $O(1)$ time
scale, and when starting from such a waterbag initial state, one
would observe the instability of the $M=0$ state at $\Ucd$ instead
of $U_c$, the phase transition energy.

Before extending these findings to other types of initial
distributions in Section~\ref{sec:extended-initial-distribution},
we will first show in the next Section that the predictions of the
Vlasov equation are sufficiently precise to give the growth rate
of the instability.

\subsubsection{Theoretical estimate of the  growth rate}
\label{sec:exponential-growth}

As shown in Fig.~\ref{fig:first-peak-a0.0}(b), a typical time
scale of the system (the one of the first peak in magnetization)
shows a tendency to diverge at $\Ucd$. We show in this Section
that Vlasov equation not only predicts this divergence, but is
also quantitative in determining the time scale of the initial
exponential instability of the magnetization for $U < \Ucd$.

Let us therefore define the exponential growth rate $\lambda$ of
an initial perturbation $\delta f(\theta,p,0)$ of a Vlasov
stationary unstable state as follows
\begin{equation}
\norm{\delta f(\theta,p,t)}\sim \exp(\lambda t)\ \norm{\delta
f(\theta,p,0)}\quad.
\end{equation}
The perturbation we consider is around a homogeneous state, whose
density function~$f$ depends only on the variable~$p$. However,
the perturbation~$\delta f$ depends on all variables
$(\theta,p,t)$.

The linearized equation that governs the time evolution of the
perturbation~$\delta f$, for vanishing magnetization,  is
\begin{equation}
\label{eq:vlasov_lin} \frac{\partial\delta f}{\partial
t}+p\frac{\partial \delta f}{\partial \theta} -\left( \sin
\theta\, M_x[\delta f] -\cos \theta\, M_y[\delta f] \right)
\frac{\partial f}{\partial p}=0\quad.
\end{equation}
Using the development in Fourier series of the perturbation
$\delta f$ given in Eq.~(\ref{dvlpfourierdeltaf}), we obtain the
following equations for each Fourier component
\begin{eqnarray}
\label{eq:fourier2}
\frac{\partial c_n}{\partial t} & = & -n p s_n \hskip 6truecm \forall n>1\\
\frac{\partial s_n}{\partial t} & = & \ \ \  n p c_n \hskip 6truecm \forall n>1 \\
\frac{\partial c_1}{\partial t} & = & - p s_1-\frac{1}{4}
\frac{\partial f}{\partial p} \int \rmd u\ s_1(u,t) \label{eq:fourier2c}\\
\frac{\partial s_1}{\partial t} & = &  \ \ \  p c_1+\frac{1}{4}
\frac{\partial f}{\partial p} \int \rmd u\
c_1(u,t)\quad.\label{eq:fourier2d}
\end{eqnarray}
One can easily show that the Fourier components with $n>1$ cannot
be unstable.  Indeed, by introducing the quantity $v_n=c_n+ i
s_n$, it is straightforward to show that its time derivative is
$v_n^{\prime}=i p n v_n$. Hence, the generalized eigenvalues are
all pure imaginary numbers, $inp$  (with $p\in\mathbb{R}$), and
the corresponding eigenvectors are Dirac delta functions.

Let us then concentrate  on the $n=1$ components for waterbag
initial conditions.  Using expressions~(\ref{eq:waterbag_deri})
for the derivative with respect to $p$ of the
distribution~(\ref{eq:waterbag}), Eqs.~(\ref{eq:fourier2c})
and~(\ref{eq:fourier2d}) become
\begin{eqnarray}
\label{eq:c1ands1} \frac{\partial c_1}{\partial t}&=&
-ps_1-\frac{1}{8\bar{p}}\left[
\delta(p+\bar{p})-\delta(p-\bar{p})\right]\int du\, s_1(u,t) \\
\frac{\partial s_1}{\partial t} &=& \ \ \
pc_1+\frac{1}{8\bar{p}}\left[
\delta(p+\bar{p})-\delta(p-\bar{p})\right]\int du\, c_1(u,t)
\quad.
\end{eqnarray}
To solve this infinite dimensional eigenvalue problem, we restrict
to functions $c_1$ and $s_1$ that are linear combinations of
$\delta(p+\bar{p})$ and $\delta(p-\bar{p})$. This yields a four
dimensional problem, whose eigenvalues can be calculated
explicitly as functions of~$\bar{p}$, and therefore as functions
of the energy per particle $U$.

\begin{figure}[htbp]
  \centering
    \includegraphics[width=7cm]{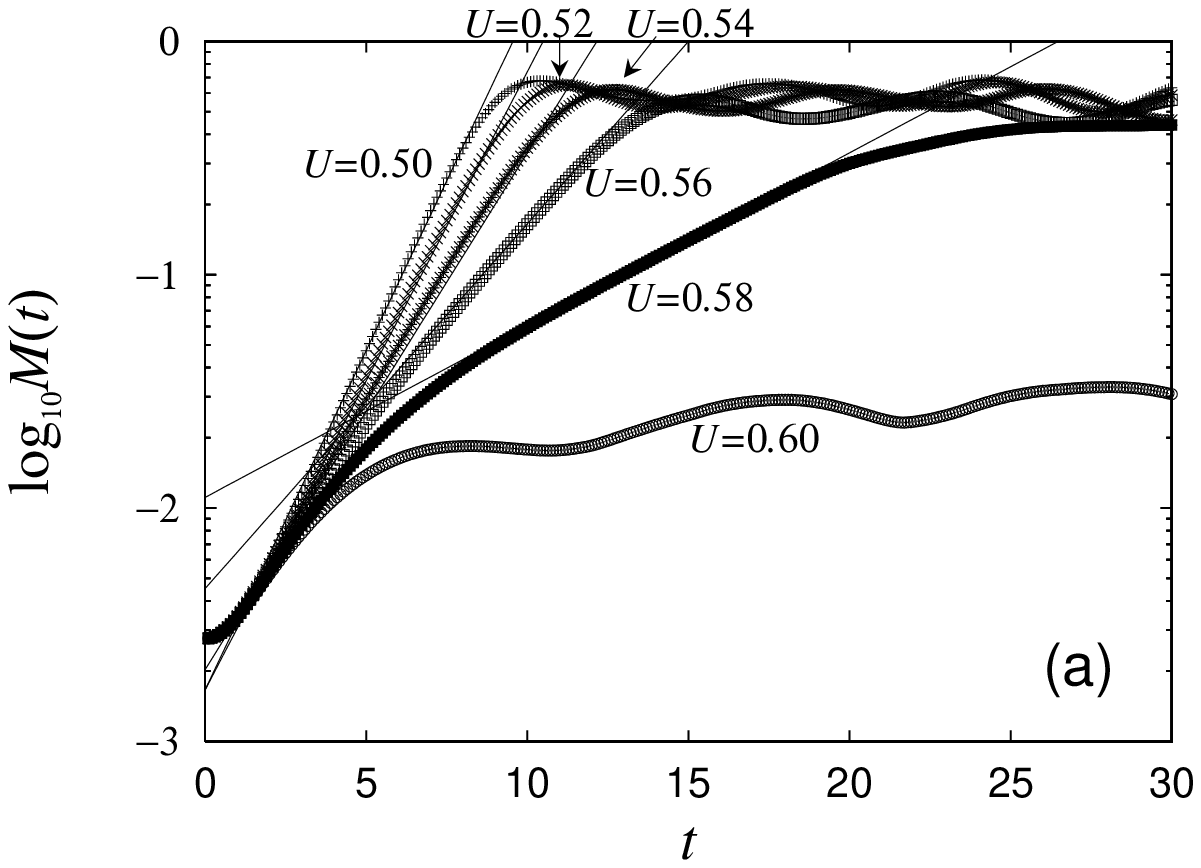}
    \includegraphics[width=7cm]{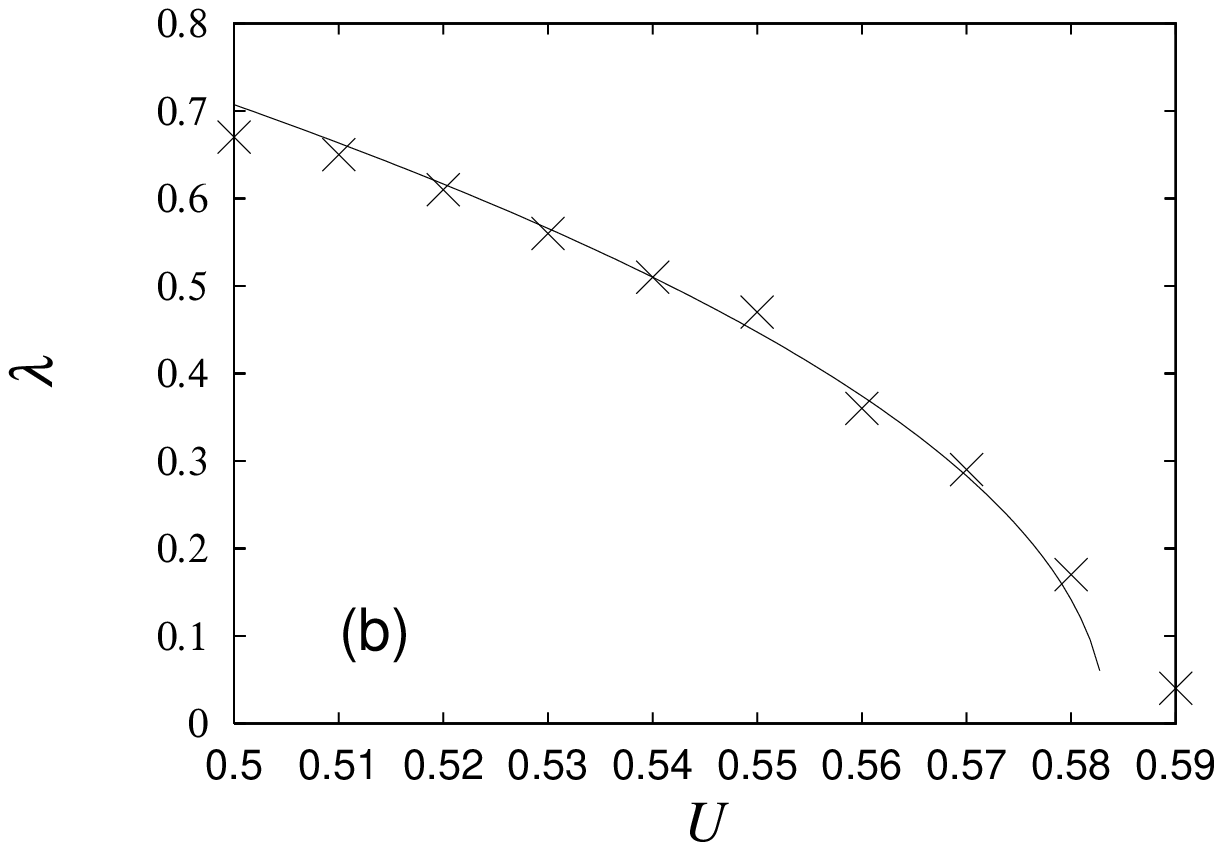}
  \caption{
    (a) Semi-log plot of the magnetization $M(t)$ for $N=10^5$
    particles, the number of samples being $20$.
    The values of $U$ are from $0.50$ to $0.60$
    from top to bottom with $0.02$ step size.
    The exponential growth rates of $M(t)$, $\lambda$,
    are estimated from the fitting solid lines.
    (b) Comparison of the theoretical (solid curve)
    and numerical (crosses) growth rate~$\lambda$.}
  \label{fig:semi-log-M}
\end{figure}

One finds that above $\Ucd=7/12$ all eigenvalues are purely
imaginary, indicating that the Vlasov equation is linearly stable
for such perturbations, in full agreement with the stability
analysis developed in Sec.~\ref{Applicationsnonlinearstability}.
Below $\Ucd$, the largest eigenvalue $\lambda$ that controls the
growth rate is
\begin{equation}
  \label{eq:1st-ev}
  \lambda = \sqrt{6(\Ucd-U)}\quad.
\end{equation}
The exponential growth of perturbations of the initial
distribution implies an exponential growth of
magnetization~$M(t)$.  This is indeed confirmed in
Fig.~\ref{fig:semi-log-M}(a) where the magnetization is plotted
versus time in semi-logarithmic scale. The comparison of the
theoretical estimate~(\ref{eq:1st-ev}) of the growth rate with the
numerical results, reported in Fig.~\ref{fig:semi-log-M}(b), shows
a very good agreement. Moreover, the time scale $1/\lambda$
diverges at $\Ucd$.

\subsection{Extended initial distribution}
\label{sec:extended-initial-distribution}

We present in this Section the numerical results for the mixed
distributions in formula ~(\ref{eq:fa}) using the indicators
introduced in Section~\ref{sec:first-peak}.

\begin{figure}[htbp]
  \centering
    \includegraphics[width=4.7cm]{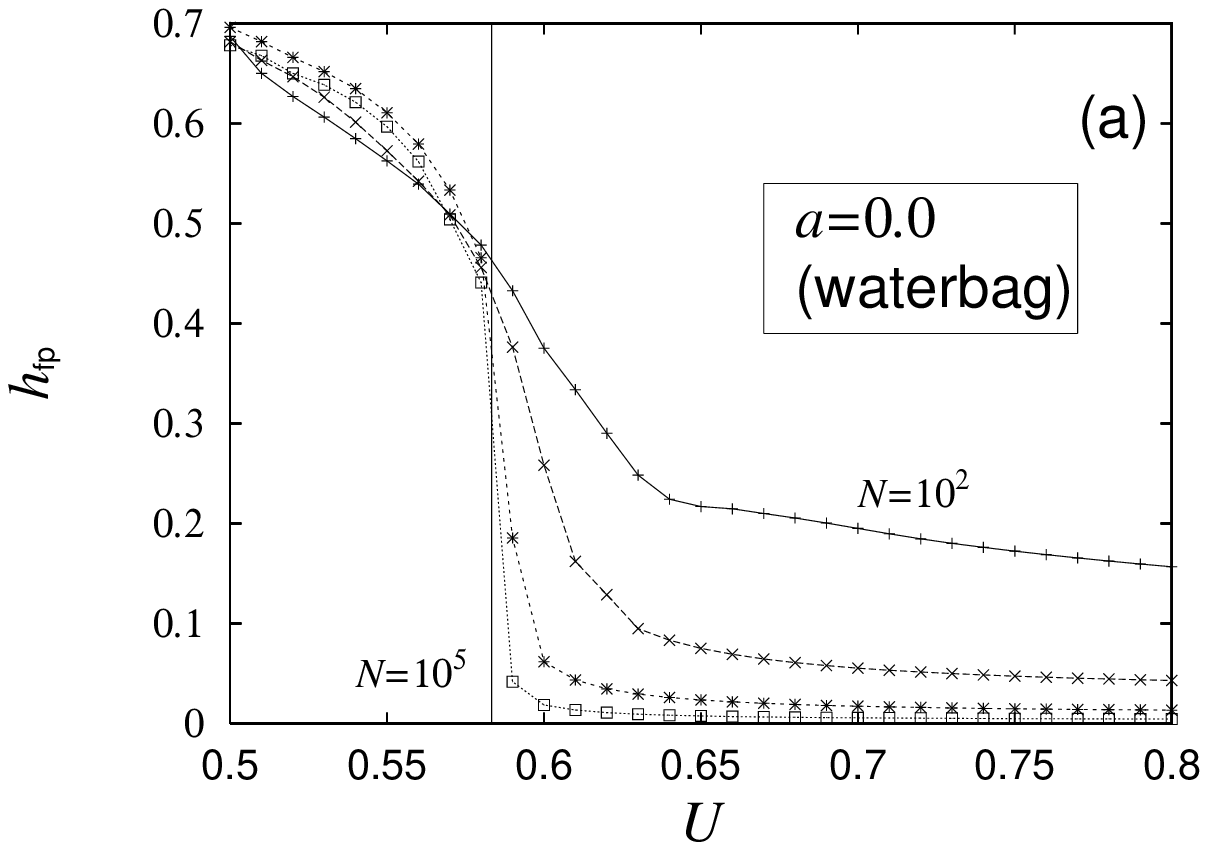}
    \includegraphics[width=4.7cm]{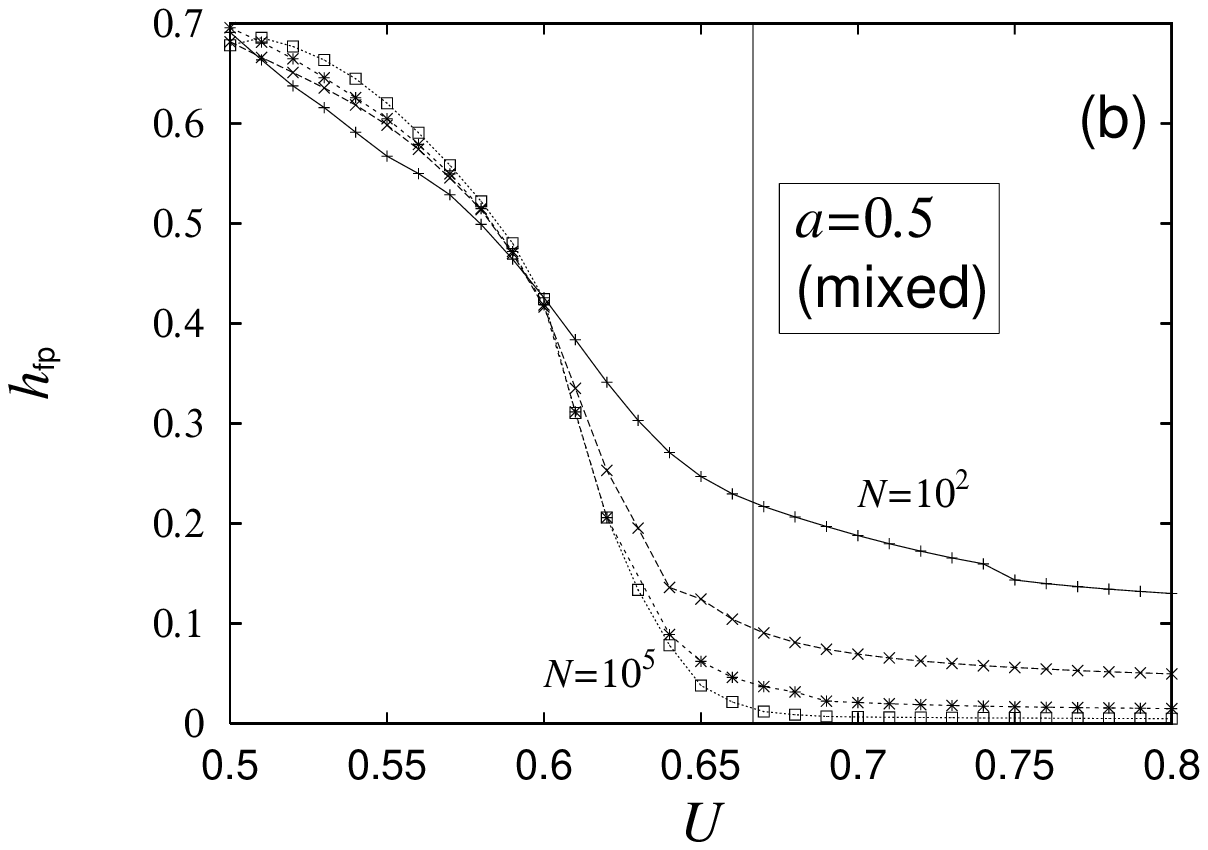}
    \includegraphics[width=4.7cm]{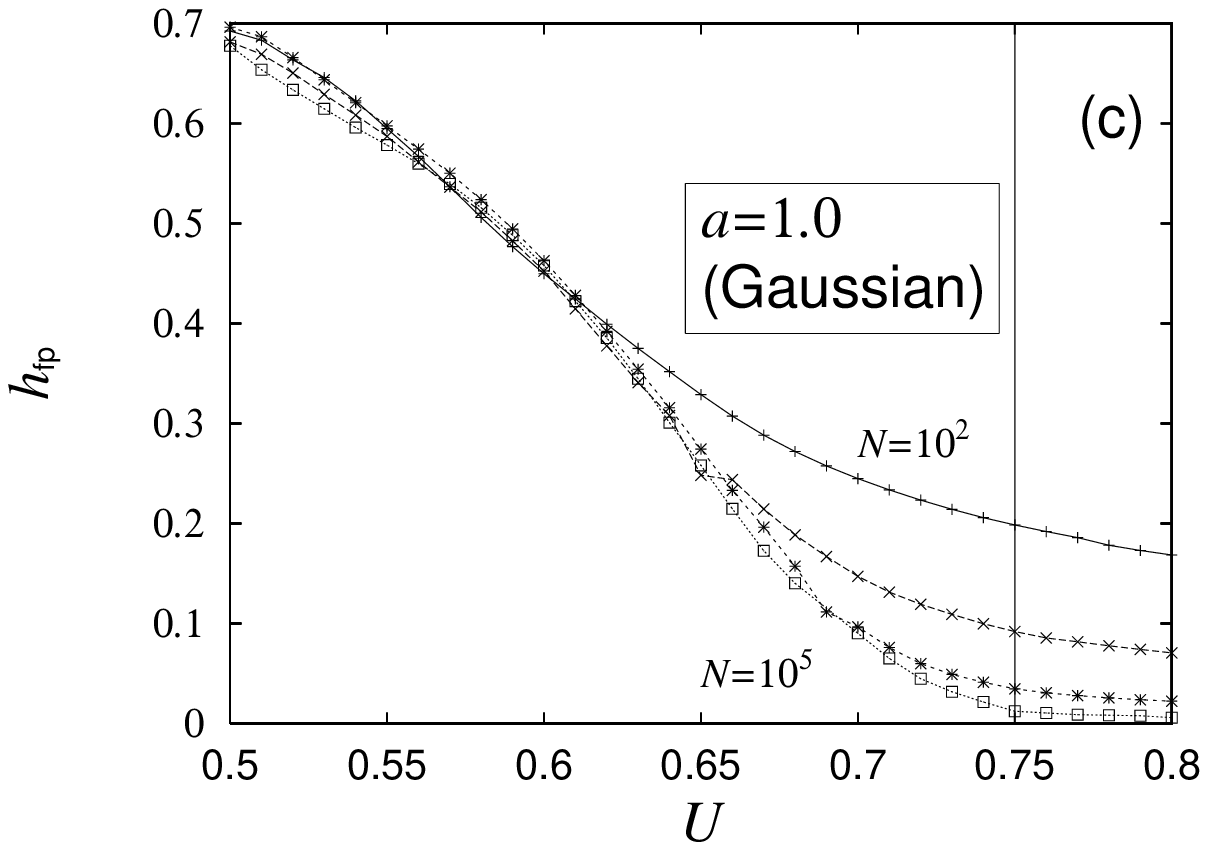}
  \caption{
    The first peak heights versus the energy density~$U$
    for waterbag (panel (a)),
    mixed with $a=0.5$ (panel (b)) and Gaussian (panel (c)) initial conditions.
    The vertical line represents the theoretical critical energy density
    $U_{c}^{\ast}(a)$, given by Eq.~(\ref{critmixed}).
    The different curves correspond to the following values of $N$:
    $10^2$, $10^3$, $10^4$ and $10^5$ from top to bottom for $U>\Ucd(a)$.
    }
  \label{fig:first-peak-height}
\end{figure}

The first peak height, shown in Fig.~\ref{fig:first-peak-height},
shows a dependence on $U$ which is perfectly consistent with the
theoretical prediction of the existence of a critical energy
density $\Ucd(a)$, given in formula~(\ref{critmixed}), above which
the $M=0$ state is stable.

However, the numerical results emphasize that the transition is
much less abrupt for non vanishing values of the parameter~$a$
than for $a=0$ (the waterbag case previously analyzed). The
explanation of this effect is possibly twofold. On one hand, the
stability criterion derived in Sec.~\ref{sec:stability} gives no
information on the time evolution that begins from an unstable
initial state: it may well be that the system evolves initially to
states with smaller magnetization than for $a=0$. On the other
hand, this weaker instability below $\Ucd$ may be also due to the
linear vanishing of the growth rate at $\Ucd$ characteristic of
the Gaussian initial distribution~\cite{Inagaki}, instead of the
sharper square root behaviour of the waterbag initial
distribution, as expressed by Eq.~(\ref{eq:1st-ev}).

\begin{figure}[htbp]
  \centering
    \includegraphics[width=4.7cm]{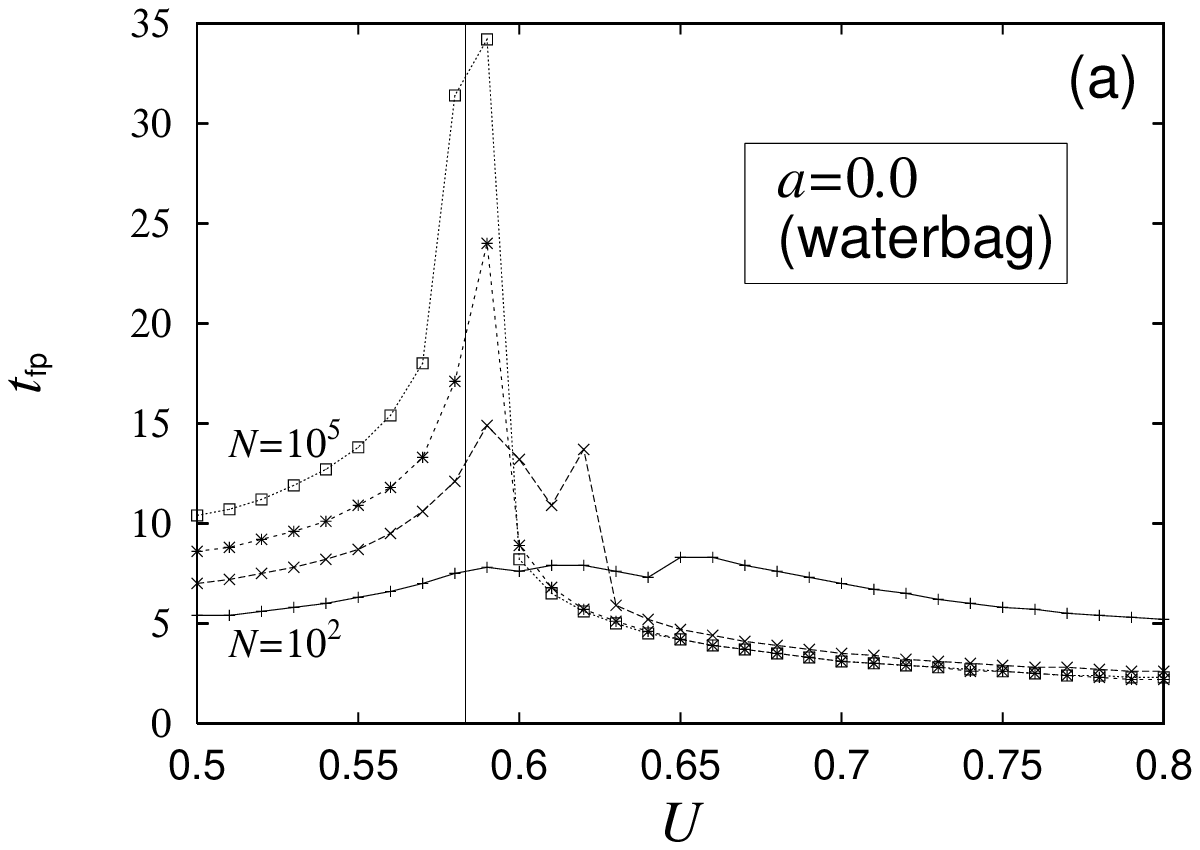}
    \includegraphics[width=4.7cm]{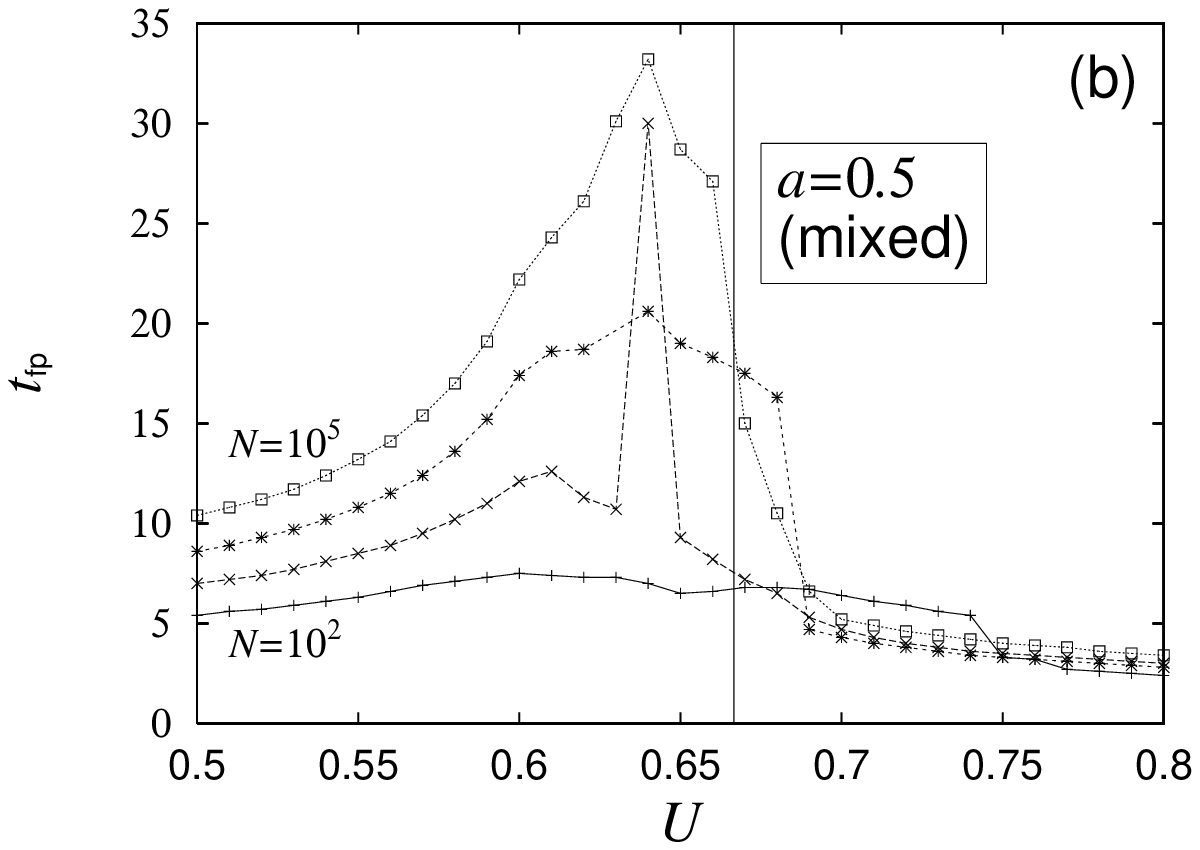}
    \includegraphics[width=4.7cm]{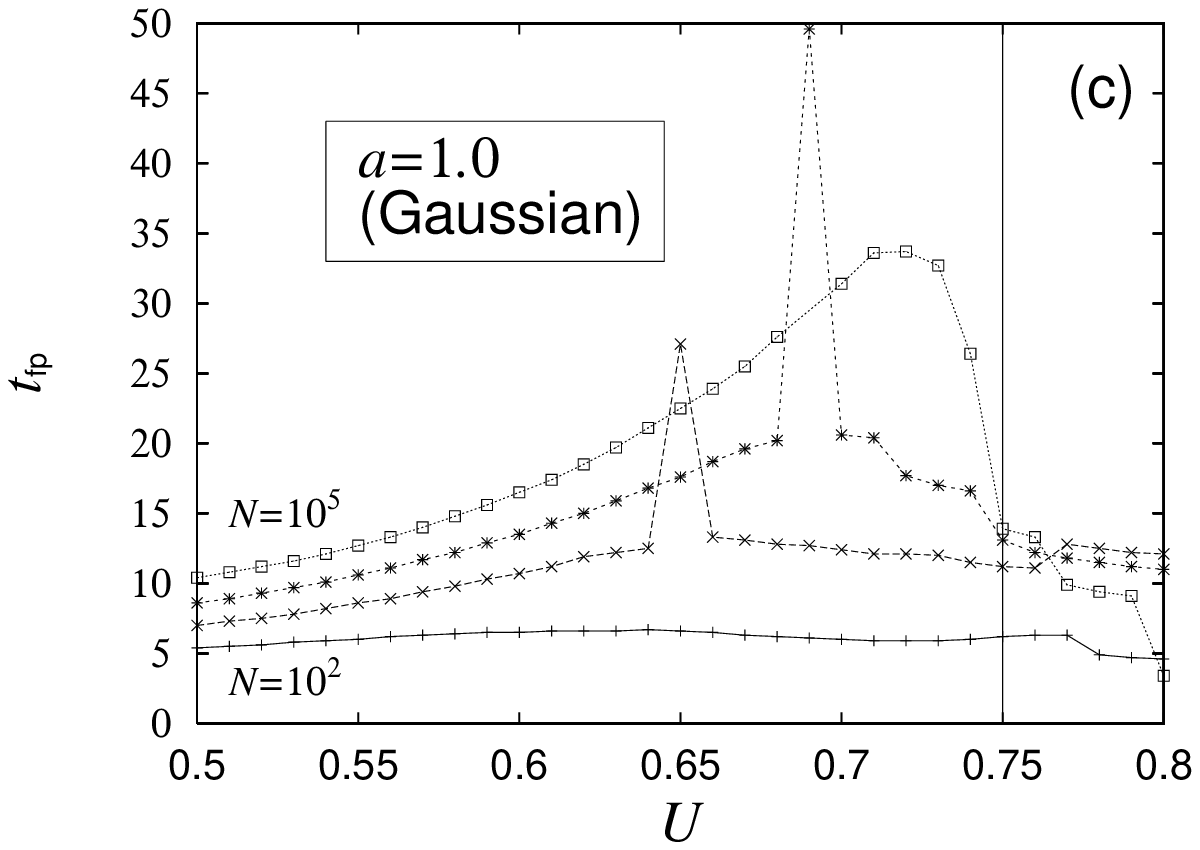}
  \caption{
    The first peak time versus the energy density~$U$
    for waterbag (panel (a)), mixed (panel (b)) and Gaussian (panel (c))
    initial conditions.
    The vertical line represents the theoretical critical energy density
    $U_{c}^{\ast}(a)$, given by Eq.~(\ref{critmixed}).
    The different curves correspond to the following values of $N$:
    $10^2$, $10^3$, $10^4$ and $10^5$ from bottom to top in $U<\Ucd(a)$.}
  \label{fig:first-peak-time}
\end{figure}

The first peak time data, presented in
Fig.~\ref{fig:first-peak-time} as a function of the energy
density~$U$, do not follow sometimes a smooth curve. This may be
due to the fact that magnetization~$M(t)$ is almost flat in the
region $U\sim\Ucd$, and hence the first peak time is strongly
affected by slight variations in the shape of the function $M(t)$.
Nevertheless, the behavior of the first peak time is qualitatively
the same as the one shown in Fig.~\ref{fig:first-peak-a0.0}(b) for
$a=0$. The same comments made there apply also to this case.

We have shown in this Section that the {\it formal stability}
criterion of Vlasov stationary states, stated in
Section~\ref{sec:stability}, allows to characterize the short time
($O(1)$ time scale) behavior of the finite $N$ dynamics of the HMF
model in the $N\to\infty$ limit. In particular, it describes the
behavior of the magnetization as a function of the energy per
particle $U$ and the presence of several different instability
thresholds depending on the detailed properties of the chosen
initial distribution. The critical energy for the instability of
the Gaussian distribution coincides with the thermodynamical phase
transition point $U_c$, as already remarked by
Inagaki~\cite{Inagaki}. Moreover, the linear stability analysis of
the Vlasov equation gives results for the growth rates of the
instability that are perfectly confirmed by numerical simulations.
Several aspects remain to be ascertained. An important one
concerns the intermediate and long time evolution of an initial
Vlasov unstable state.  In the following Section, after
introducing suitable indicators of stationarity we will follow the
time evolution of such states for the HMF model.  A further
important question is the ultimate fate of Vlasov stable $M=0$
states for the finite $N$ dynamics of the HMF model. In the next
Section, we will show that these states display indeed a much
slower ``instability'', that becomes evident on a time scale that
increases with a nontrivial power law in~$N$.  These are examples
of the {\it quasi-stationary} states that are the main object of
study of this paper.

\section{Intermediate and long time behaviours}
\label{sec:intermediate-time}

This Section is devoted to the study of the intermediate and long
time evolution of both initially unstable and stable Vlasov
stationary states.  As discussed in the previous section, the
initial evolution is well described by Vlasov dynamics. Unstable
states quickly evolve (on times of order 1), until they approach,
and are trapped, close to stationary states of the Vlasov
equation. The system then evolves slowly through these stationary
states, until it reaches Boltzmann-Gibbs statistical equilibrium.
In this section, we will construct indicators to assess if a $N$
particle state is close to a stationary state (necessary
conditions). Using these indicators, we will show that in all
cases, during the relaxation towards equilibrium, the system
always remains close to Vlasov stationary states.

\subsection{Necessary conditions of stationarity}

In order to check these features, we need to introduce necessary
conditions for stationarity, and to perform numerical tests of non
stationarity for the HMF model. The tests guarantee non
stationarity only, they do not guarantee stationarity. In order to
obtain supporting evidences of stationarity, we introduce several
indicators of non stationarity and observe whether all of them
fail. We take this fact as a good indication of stationarity.

\subsubsection{Energy distribution}
\label{sec:energy-distribution} If $f_{st}(\theta,p)$ is
stationary, so will be the single particle energy distribution,
$f_{e}(e)$. This implies the stationarity of its moments
\begin{equation}
  \label{eq:momenta-fe}
  \average{e^{j}} = \int_{0}^{\infty} e^{j} f_{e}(e)\, \rmd e \qquad,
  \qquad \text{with}\ j\in\mathbb{N}\quad.
\end{equation}
The logical implications
\begin{equation}
  f(\theta,p,t)\ \text{stationary}\Longrightarrow f_{e}(e,t)\ \text{stationary}
\Longrightarrow \average{e^{j}}(t)\
\text{stationary}\label{genstati}
\end{equation}
guarantees that the stationarity of the energy moments is a
necessary condition for the stationarity of the distribution
$f(\theta,p,t)$.

In other words, the system is not stationary if at least one of
the moments $\average{e^{j}}(t)$ is not.  The computation of the
time derivatives $\rmd\average{e^{j}}/\rmd t$ for the first four
moments will be our Test I for non-stationarity.

\subsubsection{Symmetry with respect to the spatial variable $\theta$}
\label{sec:symmetry} The previous necessary
condition~(\ref{genstati}) is valid for all Hamiltonian models. It
is possible to derive, using symmetry arguments, another necessary
condition which is specific of the HMF model. Indeed, using the
individual particle energy~(\ref{eq:individual-energy}), the
Vlasov equation~(\ref{eq:vlasov}) for a stationary distribution
can be written as
\begin{equation}
  \label{eq:stationary-condition}
  \left[ \dfracp{e}{p} \dfracp{}{\theta}
    - \dfracp{e}{\theta} \dfracp{}{p} \right]
  f_{st}(\theta,p) = 0\quad.
\end{equation}
The solution of Eq.~(\ref{eq:stationary-condition}) is constant on
the characteristic curves $e(\theta,p)=$constant. Moreover, if one
always resets to zero the phase of the magnetization vector
$\overrightarrow{M}(t)$ (i.e. $M_{y}(t)=0$), all characteristic
curves will be symmetric with respect to a sign reversal
of~$\theta$. Hence, if $f$ is stationary and the previous
condition on the phase of $\overrightarrow{M}(t)$ is respected,
then the stationary distribution obeys the following symmetry
condition
\begin{equation}
  \label{eq:symm-P}
  f_{st}(\theta,p) = f_{st}(-\theta,p), \quad \forall p~.
\end{equation}
Consequently, an asymmetry of the distribution $f(\theta,p,t)$
with respect to the spatial variable $\theta$, implies that the
system is non stationary.

It is important to notice that the symmetry with respect to sign
changes in momentum $p$, $f(\theta,p,t)=f(\theta,-p,t)$, is not a
necessary consequence of the stationarity of $f$. Two separate
characteristic curves for $e>e_{s}$, where $e_{s}$ is energy of
the separatrix, may correspond to different values of $f$ since
they are {\em not} connected.  This symmetry is therefore not
required even if the system is stationary. A violation of this
symmetry is not therefore an acceptable test of non stationarity.

\begin{figure}[htbp]
  \centering
  \subfigure[An example of asymmetric distribution]{
    \includegraphics[width=7cm]{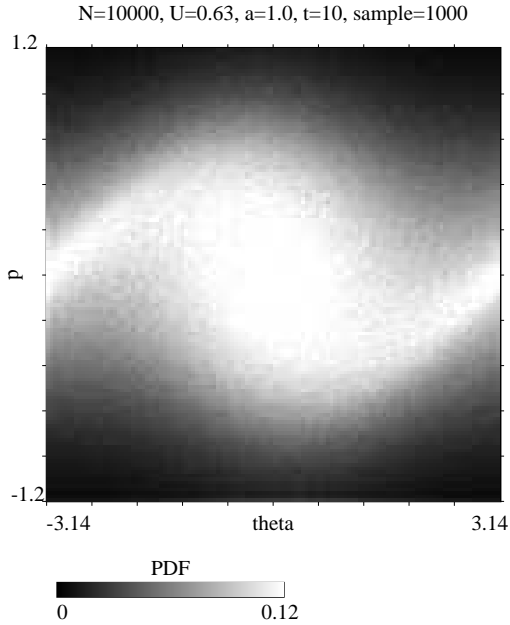}}
  \subfigure[An example of symmetric distribution]{
    \includegraphics[width=7cm]{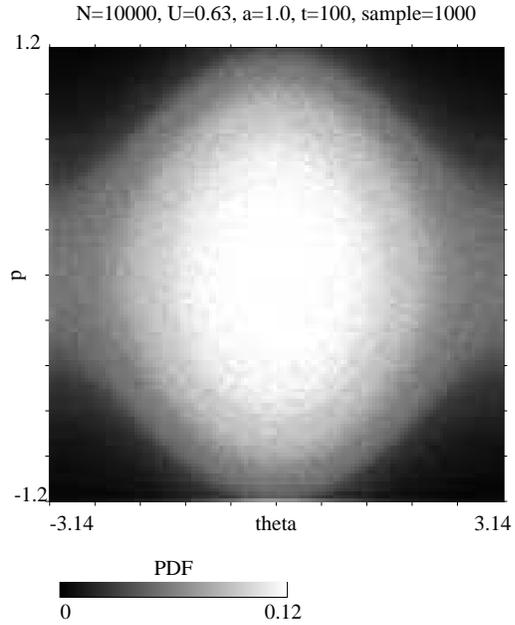}}
  \subfigure[$\tilde{f},\tilde{f}_{+}$ and $\tilde{f}_{-}$ for (a)]{
    \includegraphics[width=7cm]{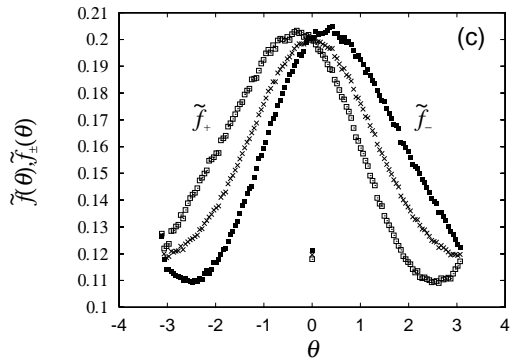}}
  \subfigure[$\tilde{f},\tilde{f}_{+}$ and $\tilde{f}_{-}$ for (b)]{
    \includegraphics[width=7cm]{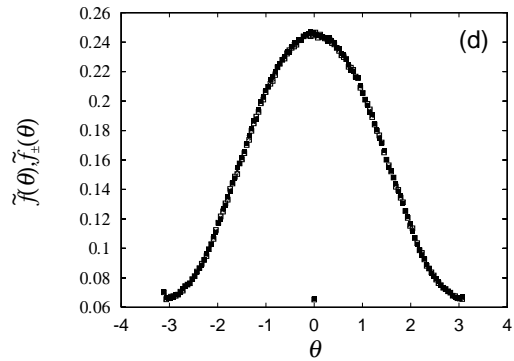}}
  \caption{
    Typical asymmetric (a) and symmetric (b) distribution
    $f(\theta,p,t)$.  Panel (c) (resp.~(d)) presents
    the marginal distributions
    $\tilde{f}(\theta,t)$ (crosses),
      $\tilde{f}_{+}(\theta,t)$ (open squares)
      and $\tilde{f}_{-}(\theta,t)$ (full squares)
    for the distributions plotted in panel (a) (resp. (b)).
    The three curves superpose almost perfectly in panel (d).
    $N=10^4$, $U=0.63$, $a=1.0$, the number of samples is $10^3$.
    Time is $t=10$ for panels (a) and (c), and $t=10^2$
    for panels (b) and (d).}
  \label{fig:asymmetry}
\end{figure}

Obtaining the distribution $f(\theta,p,t)$ numerically is,
however, not easy since it is defined on the $2$-dimensional
$\mu$-space at all fixed times $t$, and therefore a huge number of
samples would be necessary to obtain a good statistics. A trick to
easily check asymmetries in $f(\theta,p,t)$ would be to verify the
associated symmetry of the marginal distribution function
\begin{equation}
  \label{eq:integral-p}
  \tilde{f}(\theta,t) = \int f(\theta,p,t) \,\rmd p\quad .
\end{equation}
However, unfortunately, $\tilde{f}(\theta,t)$ may be symmetric
even if the symmetry (\ref{eq:symm-P}) is not satisfied.  Typical
examples of asymmetric distributions $f(\theta,p,t)$ and
corresponding symmetric marginal distributions
$\tilde{f}(\theta,t)$ are shown in Figs.~\ref{fig:asymmetry}(a)
and (c), respectively.  In order to recognize the asymmetry of
$f(\theta,p,t)$ with respect to $\theta$, one must consider
partially integrated distributions
\begin{equation}
  \label{eq:tildeP+-}
  \tilde{f}_{+}(\theta,t) = \int_{p>0} f(\theta,p,t) \,\rmd p
  \qquad\text{and}\qquad
  \tilde{f}_{-}(\theta,t) = \int_{p<0} f(\theta,p,t) \,\rmd p\quad.
\end{equation}
These latter distributions, besides being symmetric if the system
is stationary, are able to correctly detect the asymmetries of
$2D$ distributions in $(\theta,p)$, as shown in
Figs.~\ref{fig:asymmetry}(a) and (c).

We quantitatively observe the time evolution of the asymmetry by
introducing the two following quantities
\begin{equation}
  \label{eq:asymm-test}
  \begin{split}
    A_{+}(t)
    & = \int \left[ \tilde{f}_{+}(\theta,t) - \tilde{f}_{+}(-\theta,t) \right]^{2} \,\rmd \theta
    = \int \rmd \theta \left[
      \int_{p>0} \rmd p~ ( f(\theta,p,t)-f(-\theta,p,t)) \right]^{2}\quad,\\
    A_{-}(t)
    & = \int \left[ \tilde{f}_{-}(\theta,t) - \tilde{f}_{-}(-\theta,t) \right]^{2} \,\rmd \theta
    = \int \rmd \theta \left[
      \int_{p<0} \rmd p~ ( f(\theta,p,t)-f(-\theta,p,t)) \right]^{2}\quad.
  \end{split}
\end{equation}
Both $A_+$ and $A_-$ exactly vanish when the distribution is
symmetric, but they are affected by numerical errors and finite
$N$ effects when estimated from numerical simulations. In order to
reduce such effects, we first define the quantity
\begin{equation}
  \label{eq:asymm-test2}
  A(t) = \int \rmd \theta \left[ \int \rmd p~ ( f(\theta,p,t)-f(-\theta,-p,t))
  \right]^{2}.
\end{equation}
For initial conditions that are symmetric under the transformation
$(\theta,p)\to(-\theta,-p)$, since this symmetry is conserved
during the time evolution, $A(t)=0$ for both stationary or non
stationary distributions.  Of course, also this quantity is
affected by numerical errors. To factor out such errors, we
introduce the quantities $A_{+}(t)/A(t)$ and $A_{-}(t)/A(t)$,
which are large for asymmetric distributions similar to those in
Fig.~\ref{fig:asymmetry}(a), and are instead $O(1)$ for symmetric
distributions (actual values taken for the cases shown
Fig.~\ref{fig:asymmetry} are $160$ for Fig.~\ref{fig:asymmetry}(a)
and $0.4$ for Fig.~\ref{fig:asymmetry}(b)).  Large values of the
quantities~$A_{\pm}/A$ will be our Test~II for non stationarity.

\subsection{Non-stationarity test}
\label{sec:test-energy}

\begin{figure}
  \includegraphics[width=7.5cm]{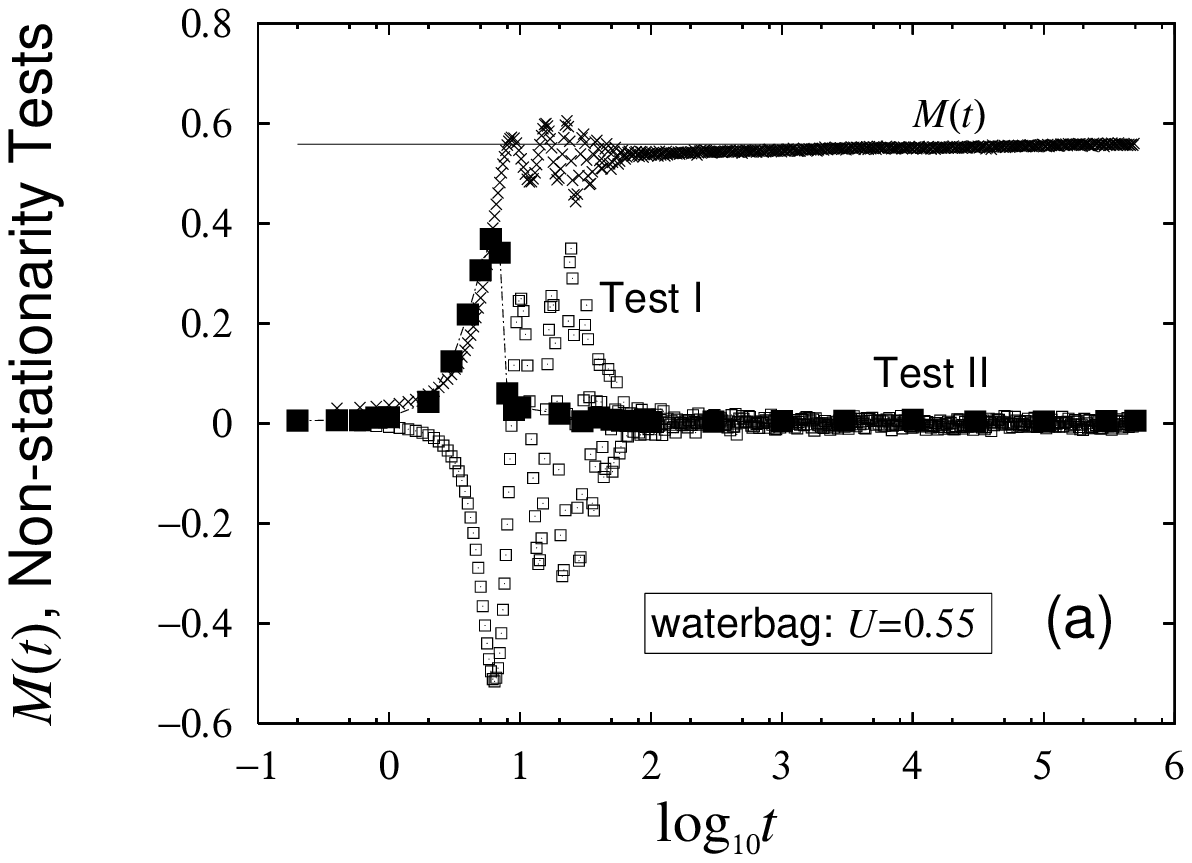}
  \includegraphics[width=7.5cm]{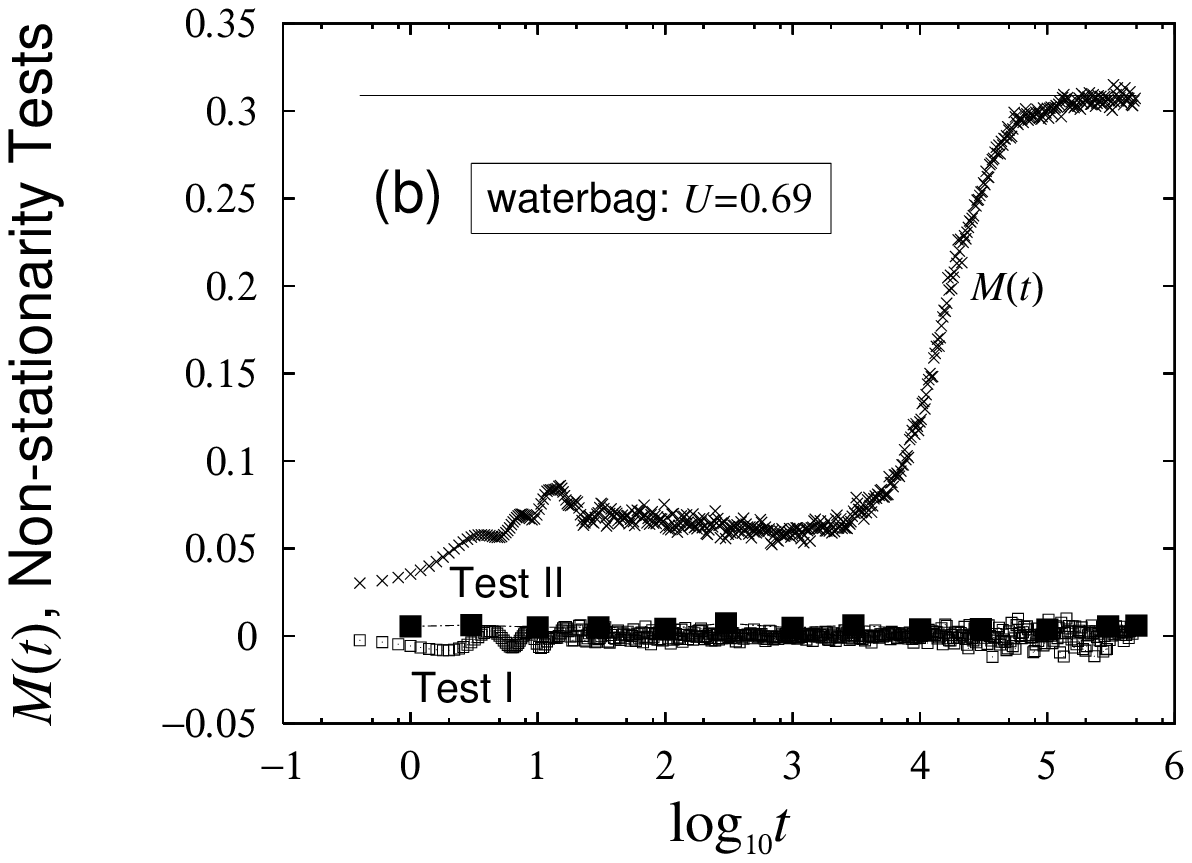}
  \caption{
    Non stationarity tests for a waterbag initial condition with
    $N=10^3$ particles and~$10^2$ samples.
    Note the logarithmic scale for the time variable.
    Panels~(a) (resp.~(b)) presents the time evolution of the
    magnetization $M(t)$ (crosses), Test I $\rmd\average{e}/\rmd t$
    (open squares) and Test II $A_{+}(t)/A(t)$ (full squares)
    in the unstable case $U=0.55$ (resp. stable case  $U=0.69$).
    The quantities $\rmd\average{e^j}/\rmd t$ (resp. $A_{\pm}(t)/A(t)$)
    are multiplied by a factor $10$ (resp. $10^{-2}$) for graphical purposes.
    The horizontal line represents the canonical value of~$M$.
  \label{fig:Test-a0.0}}
\end{figure}

Now that we have defined these two indicators, let us check non
stationarity by considering both Vlasov unstable and Vlasov stable
initial states with waterbag distribution, $a=0$ in
Eq.(\ref{eq:fa}). Typical results of Test-I and -II are shown in
Fig.~\ref{fig:Test-a0.0}.

For Vlasov unstable cases (Fig.~\ref{fig:Test-a0.0}(a)), the
quantity $\rmd\average{e^{j}}/\rmd t$ takes large values only in
the time region $2<t<80$. Similarly, the
indicator~$A_{\pm}(t)/A(t)$ is large in the time region $1<t<10$.
We have reported only $\rmd\average{e}/\rmd t$ and
$A_{+}(t)/A(t)$, but we have checked that the behaviour is totally
identical for higher moments $j=2$, 3 and 4 and for
$A_{-}(t)/A(t)$, respectively. This time region of non
stationarity perfectly coincides with the region where the
magnetization $M(t)$ rapidly grows or largely fluctuates. For
Vlasov stable cases, $\rmd\average{e^{j}}/\rmd t$ exhibits small
fluctuations only in the time interval $1<t<20$, while
$A_{\pm}(t)/A(t)$ remains small all the time. We guess that the
fluctuations in $\rmd\average{e^{j}}/\rmd t$ are due to finite
size effects and would vanish in the $N\to\infty$ limit. Note the
logarithmic scale for the horizontal time axis: the increase in
$M(t)$ in Fig.\ref{fig:Test-a0.0}(b) is much slower than the one
in Fig.\ref{fig:Test-a0.0}(a).

The behavior shown in Fig.~\ref{fig:Test-a0.0}(b) is very
important, because it shows that a variation in $M$ is compatible
with the fact that the distribution $f$ may remain stationary;
i.e. the system can relax to equilibrium (notably in the time
region $[10^4,10^5]$ in Fig.~\ref{fig:Test-a0.0}(b)) passing
through stationary states of the Vlasov equation. It is natural to
suppose in addition that these stationary states are stable.

\begin{figure}
  \includegraphics[width=7.5cm]{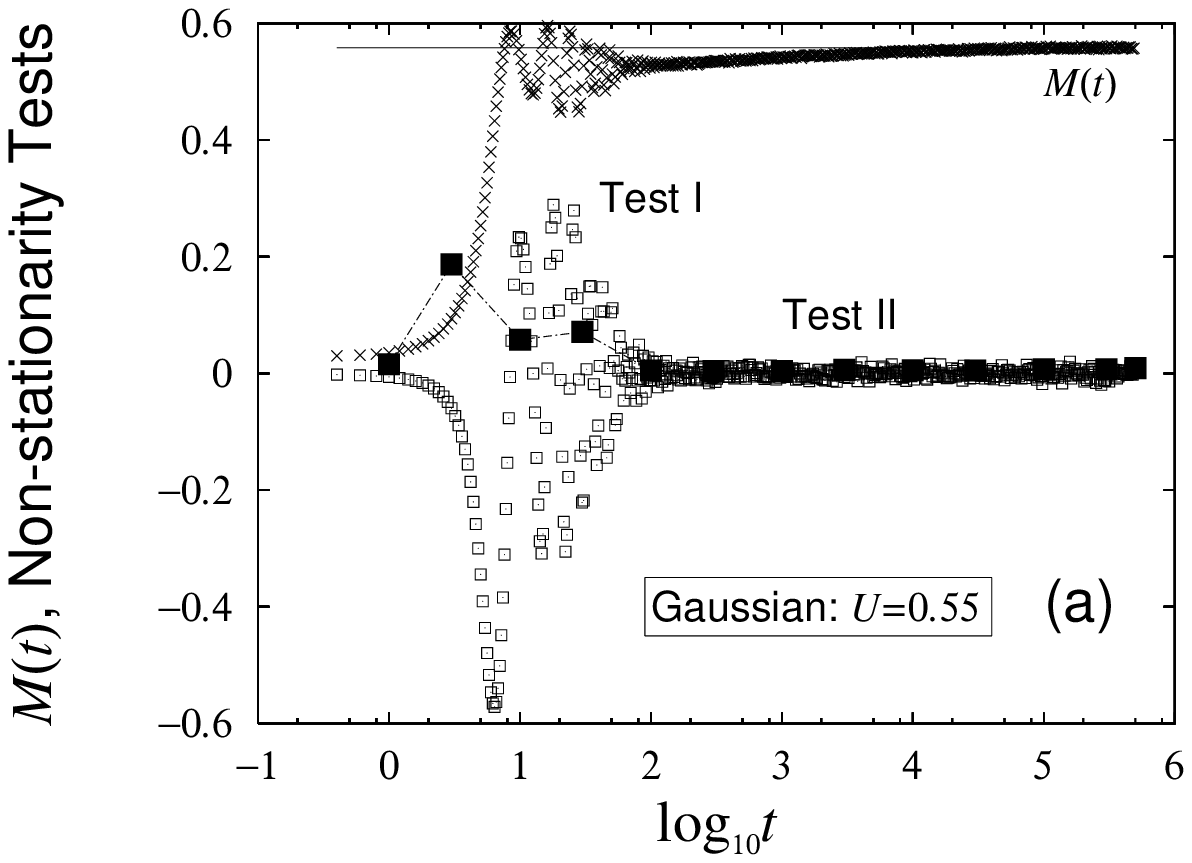}
  \includegraphics[width=7.5cm]{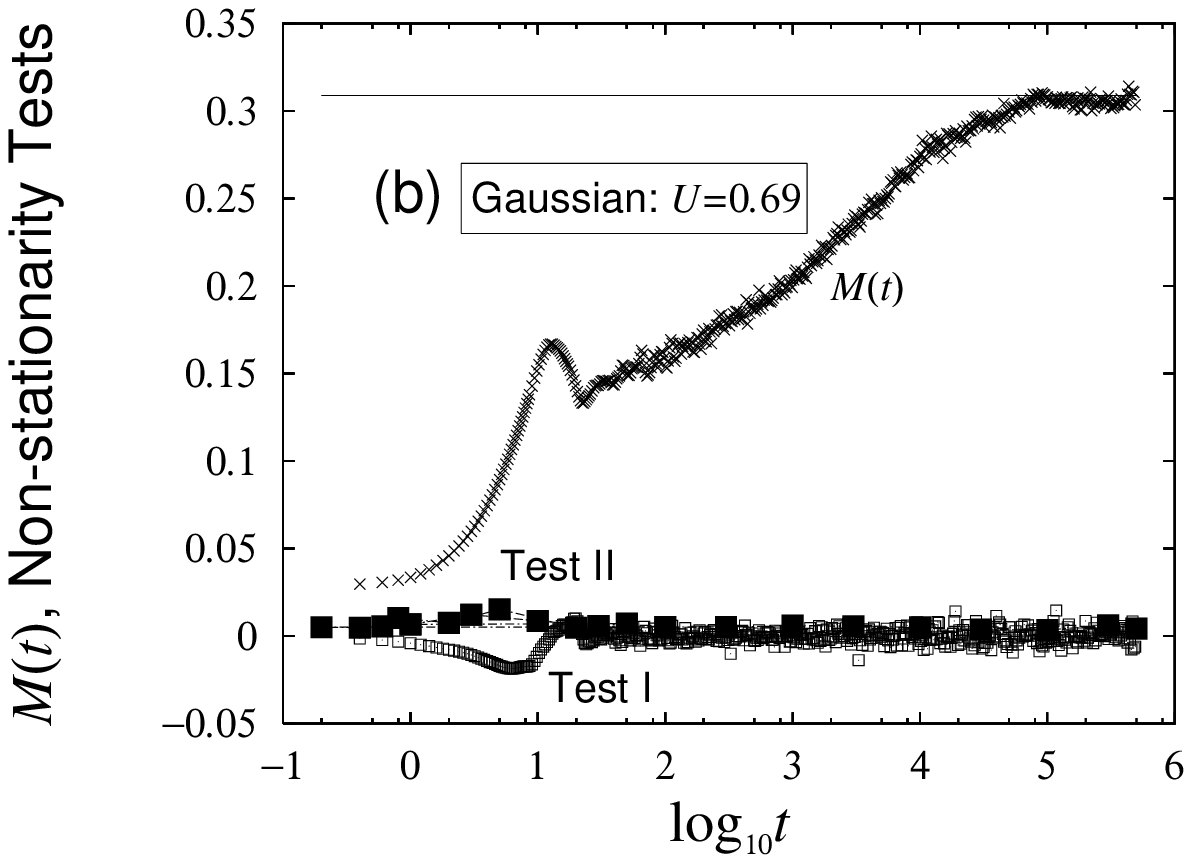}
  \caption{
    Same plots as in Figs.~\protect\ref{fig:Test-a0.0}, for a
    Gaussian initial condition.
  \label{fig:Test-a1.0}}
\end{figure}

Figure~\ref{fig:Test-a1.0} reports results for two unstable
Gaussian initial conditions, $a=1$ in Eq.(\ref{eq:fa})with
$U=0.55$ and $U=0.69$. The behavior in panel (a) is very similar
to the unstable waterbag case of Figs.~\ref{fig:Test-a0.0}(a). The
case in panel (b) exhibits a small peak around $t=5$ for both
Test~I and~II. This presumably indicates that the system is weakly
non stationary at this time, meaning that the temporal variations
of the distributions $f$ are small. It is also possible to explain
the weakness of the non stationarity for Gaussian initial
conditions with the same arguments that we used when commenting
Fig.~\ref{fig:first-peak-height}.  This state is unstable below
the critical energy $U_{c}=0.75$, but the equilibrium value
for~$M$ goes to zero when $U$ tends to $U_{c}$. Hence, the
initially unstable state may be very close to stationary stable
states, and can be therefore immediately trapped into a stable
state. We expect the non stationarity to become weaker and weaker
as $U$ approaches $U_{c}$.

\begin{figure}
  \includegraphics[width=7.5cm]{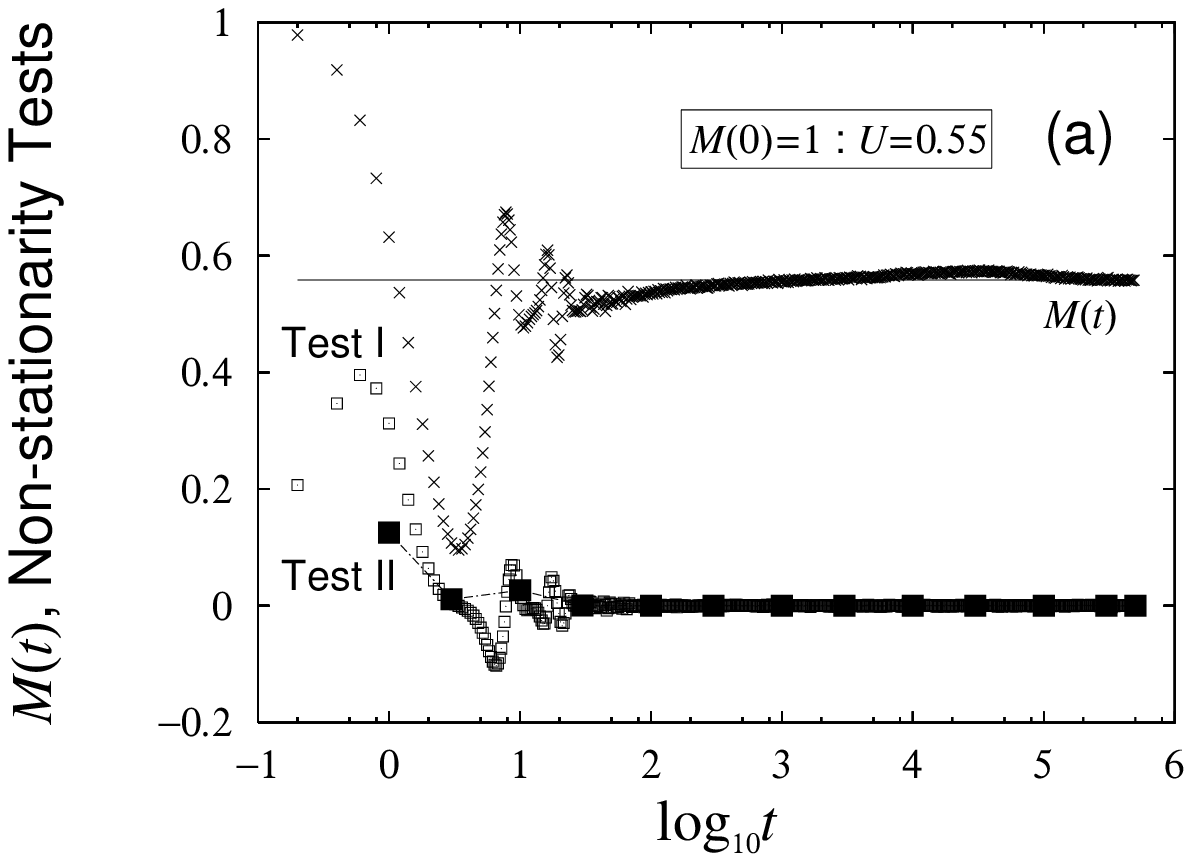}
  \includegraphics[width=7.5cm]{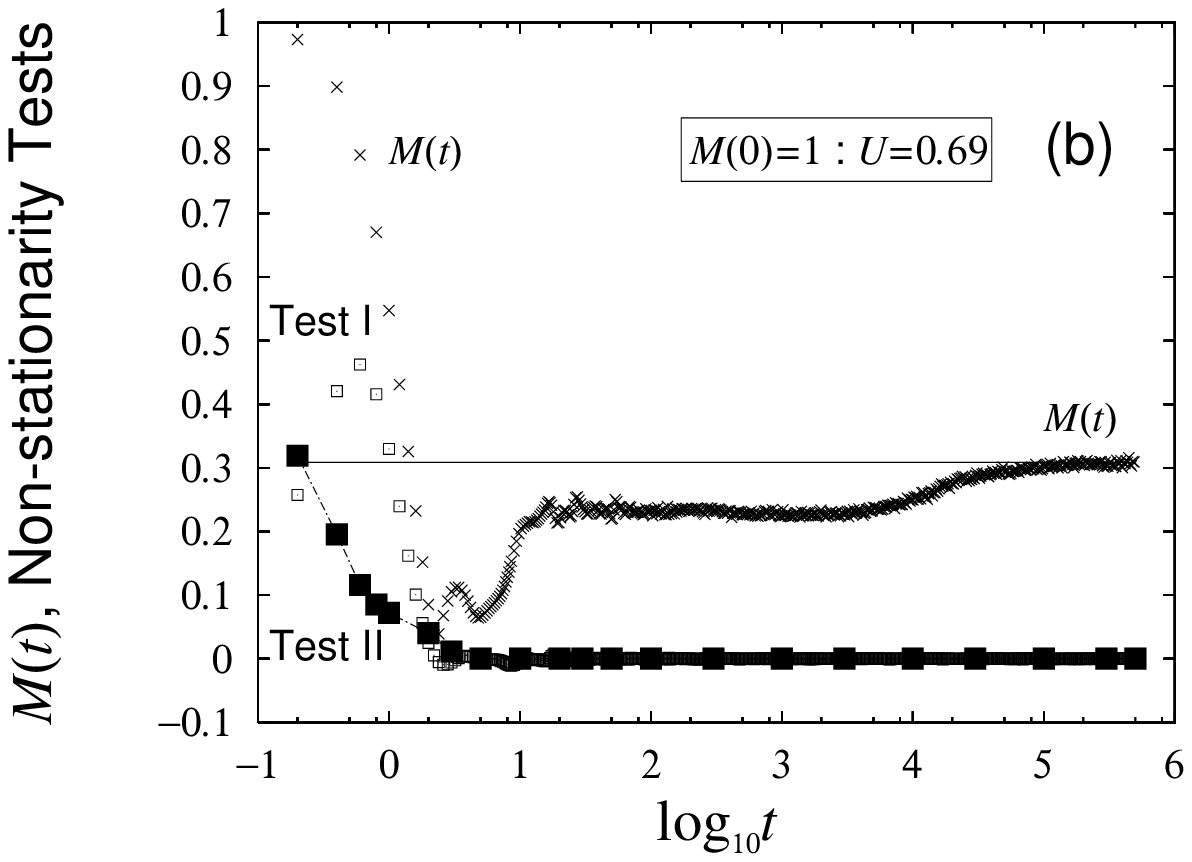}
  \caption{
    Same plots as in Figs.~\protect\ref{fig:Test-a0.0}, for
    waterbag initial conditions in momentum $p$ whereas all the phases
    $\theta_{j}$ are set at $0$, i.e. $M(0)=1$.
    The quantity $A_{+}(t)/A(t)$ is multiplied by $10^{-4}$
    for graphical purposes.
  \label{fig:Test-IC1}}
\end{figure}

The initial conditions used above correspond to stationary
solutions of the Vlasov equation. What happens with non stationary
initial conditions~? On the basis of what we have found above, we
expect that the system reaches a stationary stable state, just
like it does when starting from initially unstable stationary
states (Figs.~\ref{fig:Test-a0.0} and~\ref{fig:Test-a1.0}). To
confirm this expectation, we have prepared the initial condition
$M(0)=1$, and we have reported the results of Tests~I and~II in
Fig.~\ref{fig:Test-IC1}.  In the short time region $t<3$, both
$\rmd\average{e^j}/\rmd t$ and $A_{\pm}$ are large but they vanish
for longer times, indicating that the system reaches some
stationary state of the Vlasov equation.

Summarizing the results of this Section, we have found a strong
support to the guess that, generically, the states corresponding
to stationary stable solutions of the Vlasov equation {\em
attract} stationary unstable and non stationary solutions. The
system evolves in time passing through a series of stationary
solutions of the Vlasov equation.

\subsection{Long time behaviour }
\label{sec:long-time}

\begin{figure}[htbp]
  \centering
  \includegraphics[width=7cm]{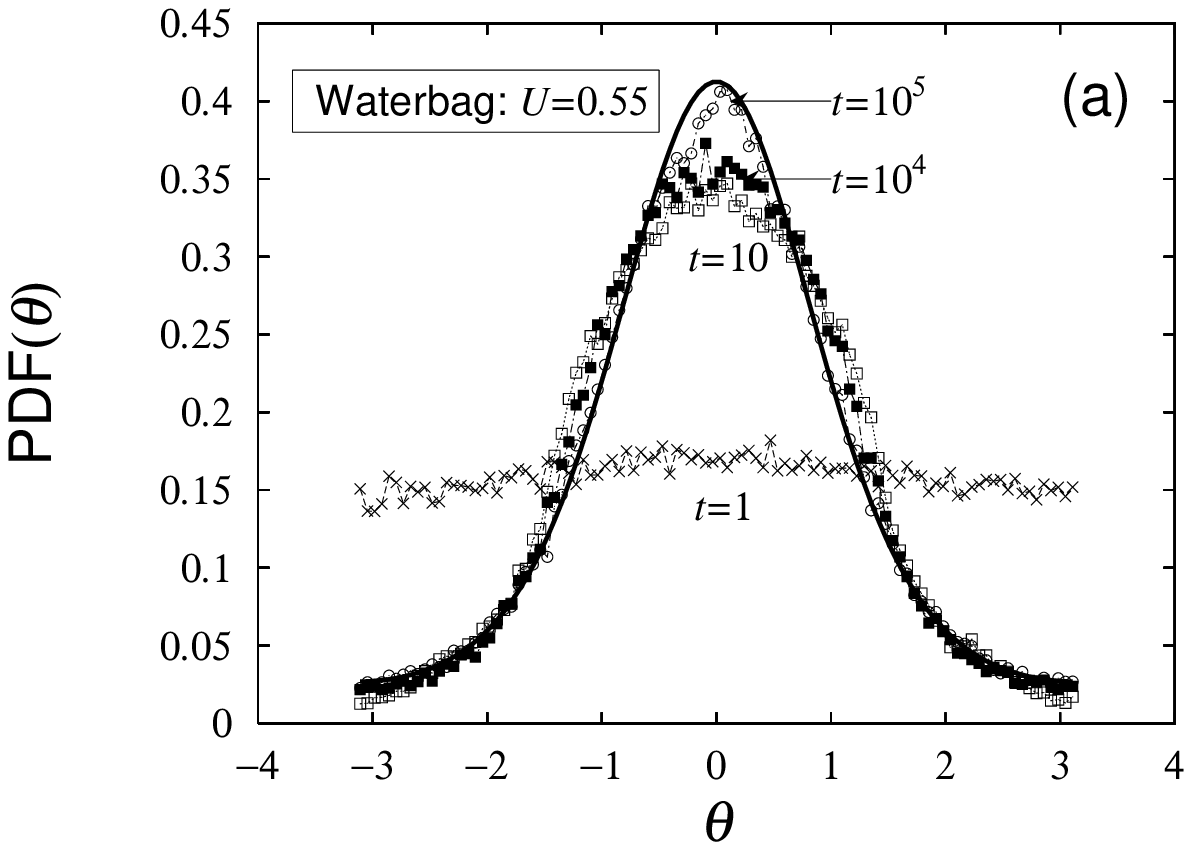}
  \includegraphics[width=7cm]{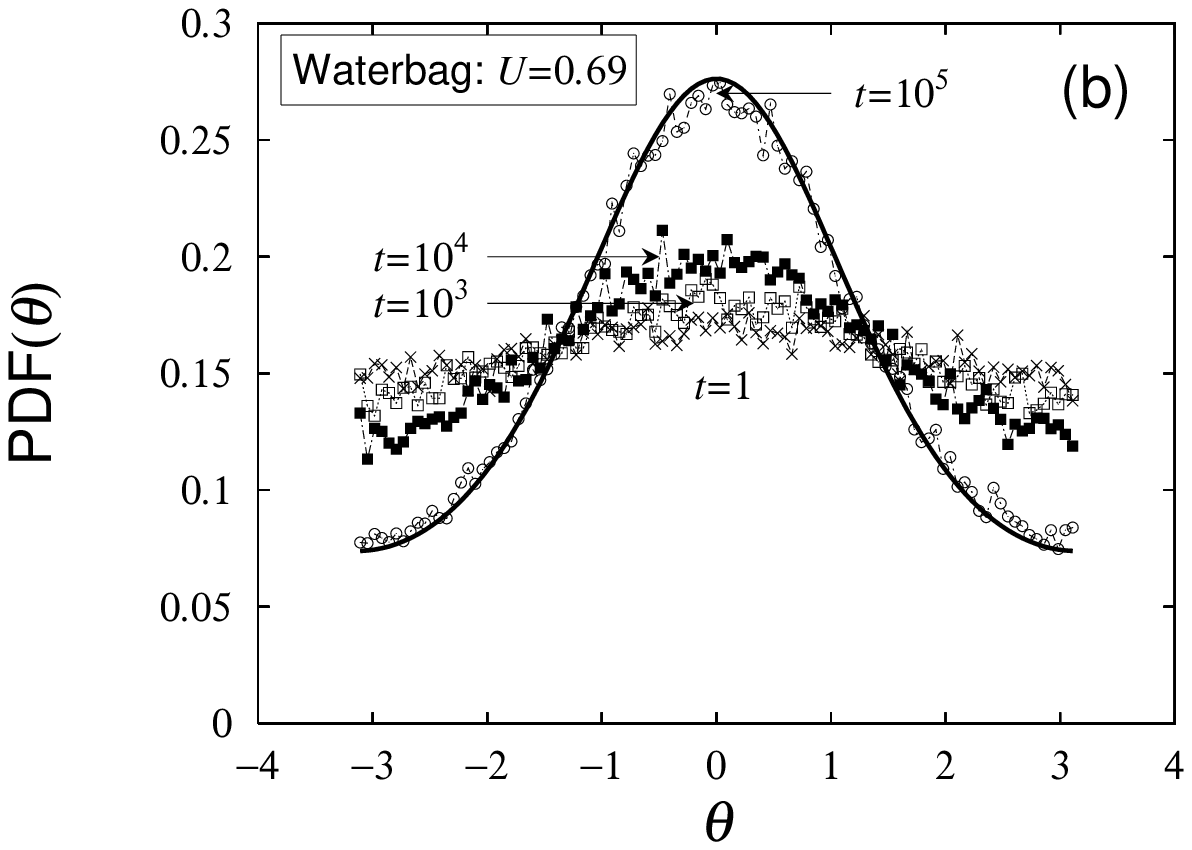}
  \caption{
    Temporal evolution of the probability distribution function
    $\tilde{f}(\theta,t)$ for a waterbag initial condition, identical to
    the one used for Fig.~\ref{fig:Test-a0.0}.
    $N=10^3$ and the number of samples is $10^2$.
    The distributions at times $t=1$ (crosses), $t=10$ (open squares),
    $t=10^{4}$ (full squares), $t=10^{5}$ (open circles),
    are reported for $U=0.55$ (a),
    and at $t=1$ (crosses), $t=10^{3}$ (open squares),
    $t=10^{4}$ (full squares), $t=10^{5}$ (open circles), for $U=0.69$ (b).
    The solid curve represents the equilibrium distribution,
    $\exp[M\cos \theta/T]/C$, where $M=0.558$,
    $T=0.412$ and $C=9.52$ for $U=0.55$ and $M=0.309$, $T=0.475$ and
    $C=6.97$ for $U=0.69$.
  \label{fig:PDFtheta-a0.0}}
\end{figure}

Let us now consider the long time behaviour of the system.  We
first discuss the behavior of the time evolution of the marginal
distribution in angle $\tilde{f}(\theta,t)=\int f(\theta,p,t) dp$,
shown in Fig.~\ref{fig:PDFtheta-a0.0}. For $U=0.55$, the initial
condition is Vlasov unstable, thus $\tilde{f}$ drastically changes
from $t=1$ to $t=10$ (see Figs.~\ref{fig:PDFtheta-a0.0}(a)
and~\ref{fig:Test-a0.0}). From $t=10$ to $t=10^4$, the
distribution is almost frozen and finally relaxes to the
equilibrium distribution at $t=10^5$. On the contrary, in the
Vlasov stable case, $U=0.69$, (see
Fig.~\ref{fig:PDFtheta-a0.0}(b)) the distribution is almost frozen
from $t=1$ up to $t=10^3$: {\it this is the crucial test of the
presence of a quasi-stationary state associated to the Vlasov
stable initial condition}. Meanwhile, the magnetization $M(t)$
remains constant, well below the equilibrium value. Finally the
distribution changes from $t=10^4$ to $10^5$, and correspondingly
the magnetization $M(t)$ increases to reach equilibrium (similar
results have been obtained for Gaussian initial distributions).

\begin{figure}[htbp]
  \centering
  \includegraphics[width=7.0cm]{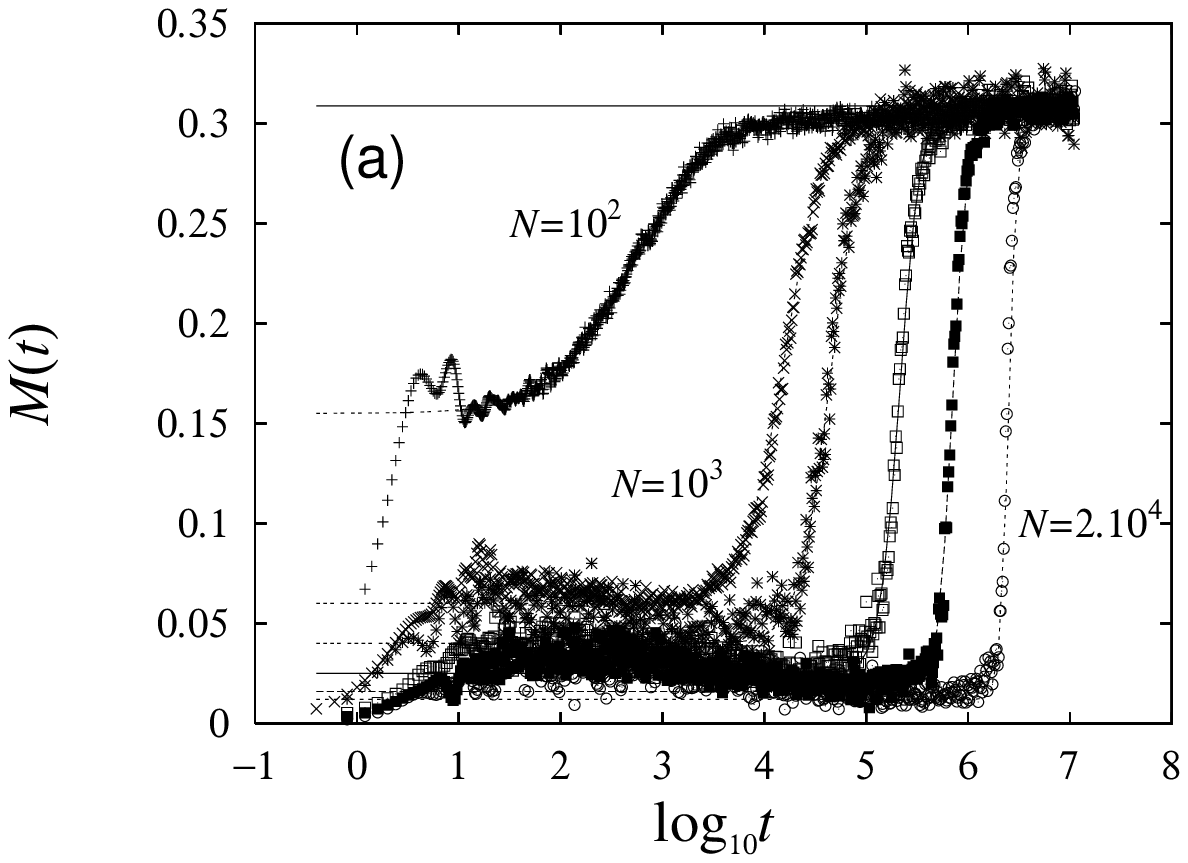}
  \includegraphics[width=7.0cm]{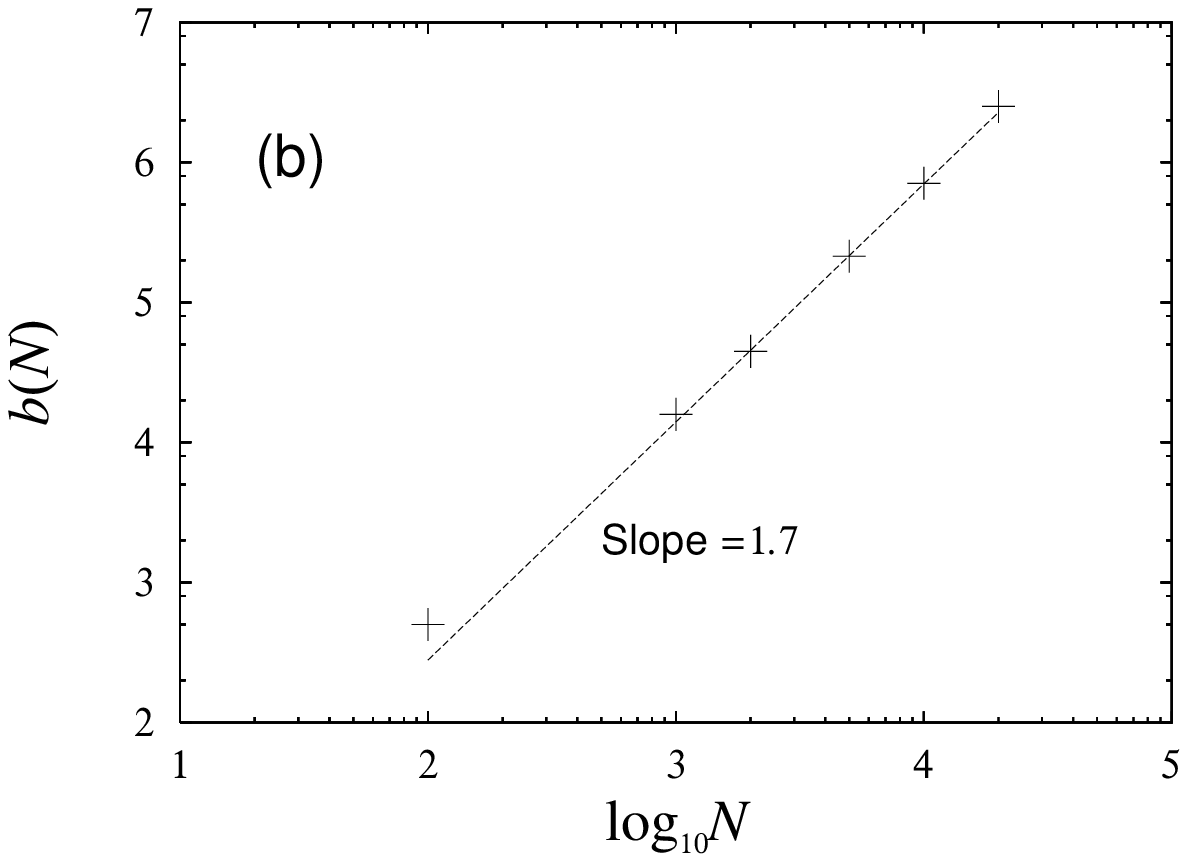}
  \caption{
    Panel (a) presents the temporal evolution of the
    magnetization~$M(t)$ for different particles numbers:
    $N=10^2(10^3)$, $10^3(10^2)$, $2.10^3(8)$, $5.10^3(8)$, $10^4(8)$
    and $2.10^4(4)$ from left to right, the number between brackets
    corresponding to the number of samples.
    The horizontal line represents the equilibrium
      value of $M$.
    Panel~(b) shows the
    logarithmic timescale $b(N)$ as a function of $N$, whereas the
    dashed line represents the law $10^{b(N)}\sim N^{1.7}$.
  \label{fig:M-apptanh}}
\end{figure}

In order to characterize the relaxation timescale with respect to
the particle number~$N$ in the case of the Vlasov stable waterbag
initial conditions ($U=0.69$), we have studied the temporal
evolution of the magnetization $M(t)$ (a preliminary study has
been already reported in Ref.~\cite{Yoshipaper}).  This can be
fitted by the function
\begin{equation}
  \label{eq:app-tanh}
  M(t) = [ 1 + \tanh ( a(N)(\log_{10}t-b(N) )) ]~c(N) + d(N),
\end{equation}
as shown in Fig.~\ref{fig:M-apptanh}(a). The parameters $c(N)$ and
$d(N)$ represent the half width between the initial and the
equilibrium levels of $M(t)$ and the initial level of $M(t)$,
respectively (we further comment about such parameters in the
Appendix).  The product $a(N)c(N)$ is the slope at
$\log_{10}t=b(N)$, i.e. $a(N)c(N)= \rmd M/\rmd
(\log_{10}t)|_{\log_{10}t=b(N)}$, and $\tau(N)=10^{b(N)}$ is the
timescale.  The most important parameter, $\tau(N)$, presented in
Fig.~\ref{fig:M-apptanh}(b) as a function of $N$, is shown to be
proportional to $N^{1.7}$. The fit is very good and excludes both
the $N$ and the $N^2$ trivial scalings. This nontrivial power law
emerges unexpectedly, since theoretical arguments as well as
previous studies of the HMF model suggest trivial divergences of
the timescale (typically as $N$) (see Ref.~\cite{HMFSpringer}).
Quite interestingly however, Zanette and
Montemurro~\cite{zanette-02} have also investigated the timescale
for the HMF model for $U=0.69$ with a waterbag initial
distribution of momenta, but with a fully ordered initial state,
i.e. $M(0)=1$. They have shown that $M(t)$ presents a minimum at
$t_{min}$ (already present in the simulation by Latora et
al~\cite{lrt2001}) which is also proportional to $N^{1.7}$.
Although our initial condition is different and hence different
scalings could be found, we consider this finding as a
confirmation of the presence of nontrivial timescales in the HMF
model. Other, apparently unrelated, nontrivial timescales have
been found, associated to the vanishing Lyapunov exponent, for
instance the well verified $1/3$ scaling~\cite{HMFSpringer}.

\begin{figure}[htbp]
  \centering
  \includegraphics[width=7.0cm]{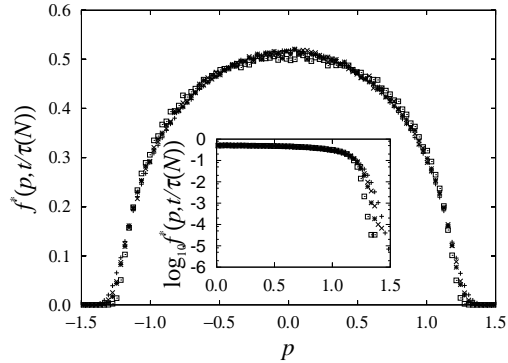}
  \caption{Temporal evolution of the momentum distribution
    $f^*(p,t/\tau(N))$ at times rescaled by $N^{1.7}$ for the Vlasov
    stable case $U=0.69$: $N=10^3$ (pluses), $N=2.10^3$ (crosses),
    $N=3.10^3$ (stars) and $N=10^4$ (open squares).
    The distributions are plotted for
    $t/N^{1.7}=0.0095$, for instance, $t=6. 10^4$ for $N=10^{4}$.
    The four curves, once time-scaled, superpose almost
    perfectly.}
  \label{fig:momentumpdf}
\end{figure}

We were led by this discovery to rescale the momentum
distributions $f^*(p,t)=\int f(\theta,p,t) d \theta$ by the
nontrivial time-scale $\tau(N)\sim N^{1.7}$, see
Fig.~\ref{fig:momentumpdf}. The superposition of the curves at
different times, in the quasi-stationary state, is quite
impressive and we consider this as a strong test of the nontrivial
timescale.

There have been recently claims that momentum distributions in
quasi-stationary states, obtained from the $M(0)=1$ initial
conditions, could be fitted with single particle distributions
inspired by Tsallis statistics~\cite{lrt2001}. In this study, we
could not obtain any reasonable fit of the momentum distribution
in Fig.~\ref{fig:momentumpdf} using several values of the Tsallis
$q$ parameter with $q>1$. Moreover, the tails of the distribution
are rather sharp, as shown by the inset in lin-log scale, and we
can numerically exclude the presence of power law tails.

It should be however mentioned that in another physical context,
i.e. in the dynamics of self-gravitating $N$-body systems confined
in an adiabatic wall, the long term time evolution has been argued
to sweep through {\it stellar polytropes}, that are peculiar
stationary solutions that maximize Tsallis entropy~\cite{taruya}.
However, Chavanis~\cite{chavanisgeneralization}, using the
minimization of functionals built on conserved quantities (called
H-functions in this context), showed that Tsallis entropy can be
considered as a particular choice among the infinitely many
possible H-functions: their maximization lead to particular
stationary solutions of the Vlasov dynamics.

The success of Tsallis statistics in describing quasi-stationary
regimes in N-body dynamics remains therefore doubtful. Further
studies should fix this puzzling issue.

It should be moreover mentioned that in connection with some aging
properties of the $M=1$ initial state, $q$-exponential have been
used to fit quite efficiently numerical data of correlation
functions~\cite{Monte,Pluchino}.

\section{Summary}
\label{sec:summry}

In this paper, we have proposed a general framework, based on
Vlasov equation, to study the relaxation to equilibrium of the HMF
model. In the short time regime, the behaviour of the system is
well described by the Vlasov equation, which is a valid
approximation for finite times. The stability of the initial
conditions is therefore adequately characterized by considering
the stability of Vlasov stationary states. We have checked the
accuracy of this criterion in comparison with finite~$N$
simulations. In addition, a simple analytical derivation of the
largest eigenvalue of the linearized Vlasov operator gives an
excellent estimate of the numerical timescale in the short time
regime of the finite~$N$ system.

In the intermediate and long time regimes, two situations have
been analyzed separately. The first one, corresponding to unstable
stationary Vlasov states, shows a complex short time evolution
before reaching some stationary stable state of the Vlasov
equation that is situated close to Boltzmann-Gibbs equilibrium.
The latter is finally reached on very long times, $\tau(N)\sim N$.
The second one, which begins from Vlasov stable stationary states,
evolves through other {\it quasi-stationary} states
that are far from equilibrium and are among
the many other stationary Vlasov states.  The final relaxation to
Boltzmann-Gibbs equilibrium takes place on times that increase
with a nontrivial power law in the number of particles
$\tau(N)\sim N^{1.7}$. Interestingly, this exponent is consistent
with the one recently reported in Ref.~\cite{zanette-02} for a
different initial condition.

We have analyzed the momentum distributions in such a {\it
quasi-stationary} regime and found that they scale properly with the
nontrivial exponent. As in~\cite{Pluchino}, we have been
unable to fit the momentum distributions using $q$-exponentials, as it
had instead been done for other initial states in Ref.~\cite{lrt2001}.
This is a negative indication that {\it quasi-stationary} states can
always be described by Tsallis statistics.

An analogy between the slow time evolution appearing in {\it
quasi-stationary} states and the aging phenomenon in glasses and
spin-glasses has been proposed in Ref.~\cite{Monte} and further
analyzed in Ref.~\cite{Pluchino}.  However, none of these studies
has been performed using stationary states of the Vlasov equation
as initial conditions. It would be extremely interesting to repeat
the study of correlation functions and of other glassy behavior
indicators for such states, as already partially done in
Ref.~\cite{Yoshipaper}, where non stationary stretched exponential
correlation functions have been found.

After completing this paper, we became aware of a further, and
more complete, study by Choi and Choi \cite{Choi} of the linear
stability of the homogeneous state in both the canonical
(Fokker-Planck) and the microcanonical (Vlasov) ensemble. However,
this paper does not give any answer to the question of the
instabilities caused by the finiteness of the number of particles.

\begin{ack}
We warmly thank F. Baldovin, A. Campa, P-H. Chavanis, J-P.
Eckman, Y. Elskens, D. Escande, A. Giansanti, V. Latora, M.
Nauenberg, A. Rapisarda, F.  Tamarit, A. Taruya, C. Tsallis, R.
Vallejos, C. Villani for helpful discussions. We also thank P-H.
Chavanis, M. Nauenberg and C. Tsallis for their critical reading
of the first version of this manuscript. This work has been
partially
  supported by EU contract No.  HPRN-CT-1999-00163 (LOCNET network),
  the French Minist{\`e}re de la Recherche grant ACI jeune chercheur-2001
  N$^\circ$ 21-31, and the R{\'e}gion Rh{\^o}ne-Alpes for the fellowship N$^\circ$
  01-009261-01. This work is also part of the contract COFIN00 on {\it
    Chaos and localization in classical and quantum mechanics} and of
  the FIRB n. RBNE01CW3M\_01 project on synchronization. F.B. is
  supported by E.U. Network {\em Stirring and Mixing}
  (RTN2-2001-00285).
\end{ack}

\appendix
\section{Appendix: Detailed scaling of the magnetization}
\begin{figure}[htbp]
  \centering
  \includegraphics[width=7.0cm]{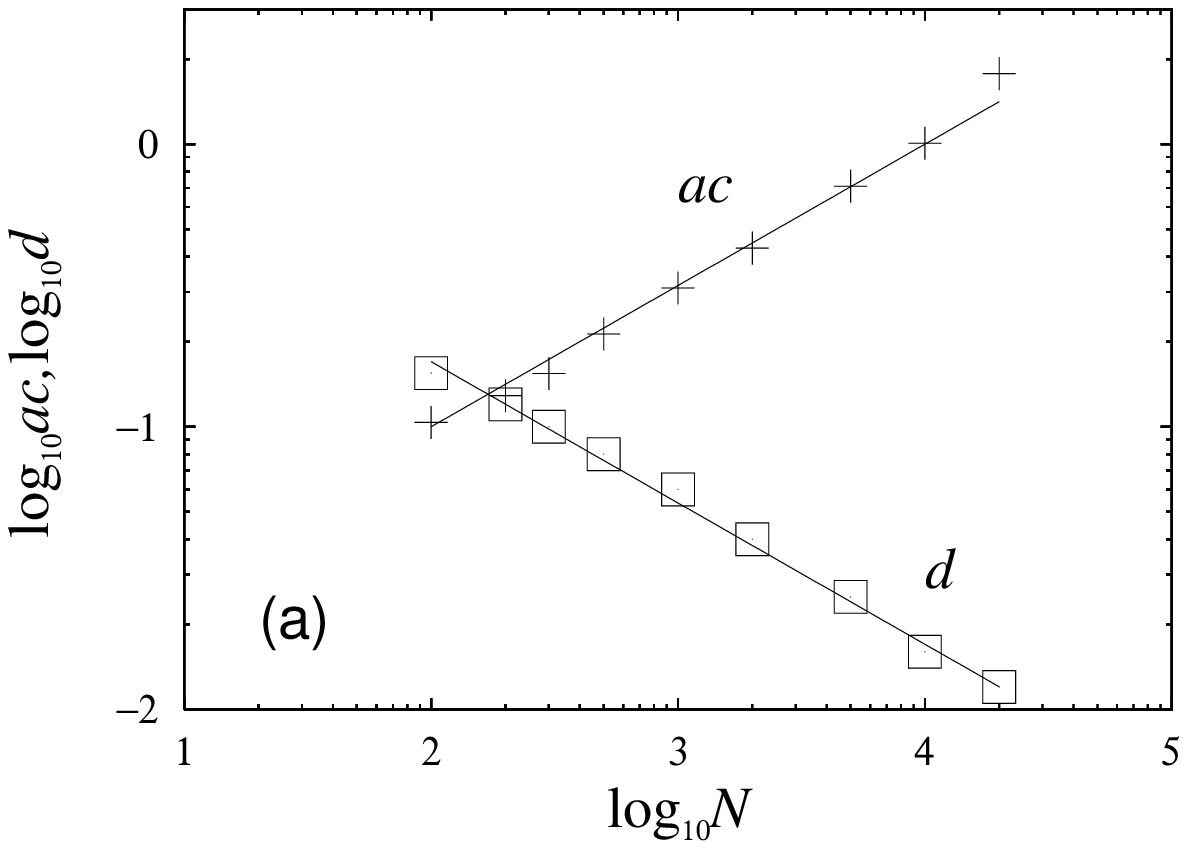}
  \includegraphics[width=7.0cm]{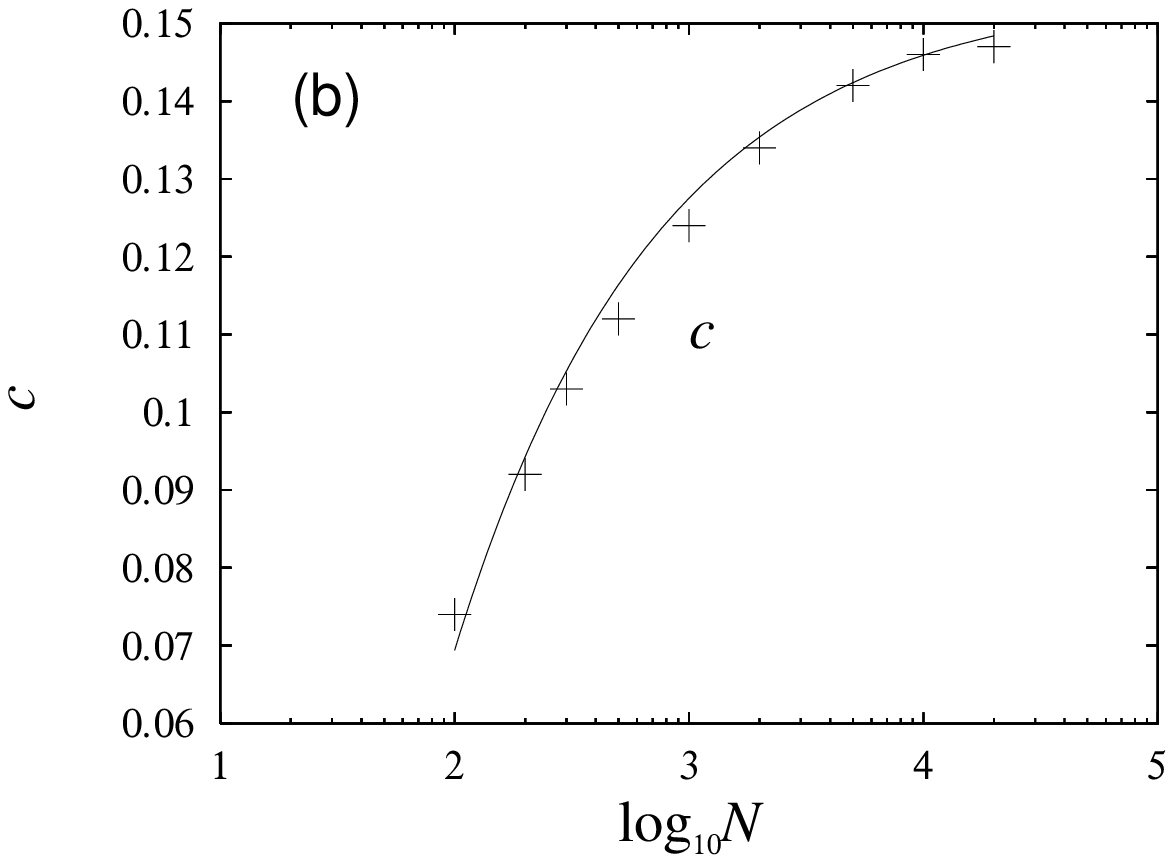}
  \caption{
    Log-log plots of parameters $a(N)c(N)$ and $d(N)$ in panel (a),
    and $c(N)$ in panel (b). The symbols represent numerical results whereas
    the curves correspond to the fitted functions, given by
    Eqs.~(\protect\ref{eq:parameter-scaling}).}
  \label{fig:M-scaling}
\end{figure}

Let us recall the fitting function we have used in
Section~\ref{sec:long-time} for the magnetization~$M(t)$
\begin{equation}
  \label{eq:app-tanh2}
  M(t) = [ 1 + \tanh ( a(N)(\log_{10}t-b(N) )) ]~c(N) + d(N)~.
\end{equation}
We have already commented about the scaling behavior of
$\tau(N)=10^{b(N)}$. According to Fig.~\ref{fig:M-scaling}, all
the parameter set scales as follows:
\begin{equation}
  \label{eq:parameter-scaling}
  a(N) = \dfrac{\sqrt{N}}{100~c(N)}, \quad
  10^{b(N)} = \dfrac{1}{9} N^{1.7}, \quad
  c(N) = \dfrac{(M_{\text{eq}}-d(N))}{2}, \quad
  d(N) = \dfrac{1.7}{\sqrt{N}},
\end{equation}
where $M_{\text{eq}}$ is the equilibrium level of $M(t)$.

\begin{figure}[htbp]
  \centering
  \includegraphics[width=7.0cm]{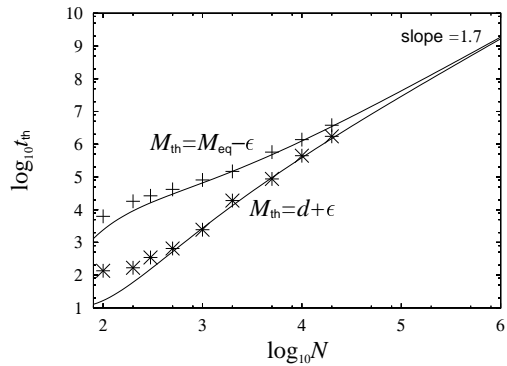}
  \caption{
    Dependence of the threshold time
    $t_{\text{th}}=t(M_{\text{th}})$ on $N$ for
    $M_{\text{th}}=d+\varepsilon$ (stars) and
    $M_{\text{th}}=M_{\text{eq}}-\varepsilon$ (crosses). Both solid
    curves are obtained from Eq.~(\protect\ref{eq:t_th}).}
  \label{fig:t-Mth}
\end{figure}

By using the scaling functions (\ref{eq:parameter-scaling}), we
can predict when $M(t)$ reaches a given threshold,
$M_{\text{th}}$, as a function of $N$.  The corresponding
threshold time, $t_{\text{th}}$, with
$M(t_{\text{th}})=M_{\text{th}}$, has the following expression
\begin{equation}
  \label{eq:t_th}
  t_{\text{th}}(M_{\text{th}}) = 10^{b} \left(
    \dfrac{M_{\text{th}}-d}{2c+d-M_{\text{th}}}
  \right)^{\dfrac{\ln 10}{2a}}
  = \dfrac{1}{9} N^{1.7} \left(
    \dfrac{M_{\text{th}} - \dfrac{1.7}{\sqrt{N}}}{M_{\text{eq}}- M_{\text{th}}}
  \right)^{\dfrac{\ln 10}{2a}}.
\end{equation}
Let us consider, for numerical purposes, two threshold times for
both threshold levels, $M_{\text{th}}=d+\varepsilon$ and
$M_{\text{eq}}-\varepsilon$. These two times roughly represent the
beginning and ending times when $M(t)$ grows towards the
equilibrium value $M_{\text{eq}}$~\cite{Yoshipaper}. Numerical
results have fluctuations, particularly in the early time region,
and hence we have defined $t_{\text{th}}(d+\varepsilon)$ and
$t_{\text{th}}(M_{\text{eq}}-\varepsilon)$ as follows:
\begin{eqnarray}
  t_{\text{th}}(d+\varepsilon)&
  =& \max \{ t\in R~ | ~  M(t) < d+\varepsilon \}\quad, \\
  t_{\text{th}}(M_{\text{eq}}-\varepsilon)
  &=& \min \{ t\in R~ | ~  M(t) > M_{\text{eq}}-\varepsilon \}\quad. \quad
\end{eqnarray}
In Fig.~\ref{fig:t-Mth}, the comparison between the numerical
results and the theoretical prediction~(\ref{eq:t_th}) for
$\varepsilon=0.0088$ shows an excellent agreement, except in the
small $N$ region. We remark that both threshold times are
approximately proportional to $N^{1.7}$ in the thermodynamic
limit. This further confirms the nontrivial scaling law.

\end{document}